\documentclass[fleqn,usenatbib]{mnras}

\usepackage{newtxtext,newtxmath}
\usepackage{graphicx}	
\usepackage{amsmath}	
\usepackage{gensymb}
\usepackage{xcolor}
\usepackage{multirow}

\usepackage{tikz,xcolor,hyperref}

\usepackage[T1]{fontenc}

\DeclareRobustCommand{\VAN}[3]{#2}
\let\VANthebibliography\thebibliography
\def\thebibliography{\DeclareRobustCommand{\VAN}[3]{##3}\VANthebibliography}

\defcitealias{Knowles_2021MNRAS}{K21}

\title[Abell 384]{Exploring the Dynamical Nature and Radio Halo Emission of Abell 384: A Combined Radio, X-ray and Optical Study}

\author[Chatterjee S. et al.]{
Swarna Chatterjee$^{1,2}$\thanks{E-mail: swarna.chatterjee@ru.ac.za}, Denisha Pillay$^{3}$, Abhirup Datta$^{1}$, Ramij Raja$^{2}$, Kenda Knowles$^{2,4}$, Majidul Rahaman$^{5}$, \newauthor S.P. Sikhosana,$^{3,6}$
\\
$^{1}$Department of Astronomy Astrophysics and Space Engineering, Indian Institute of Technology Indore, Indore, M.P., India\\
$^{2}$Centre for Radio Astronomy Techniques and Technologies, Department of Physics and Electronics, Rhodes University, Makhanda 6139, South Africa\\
$^{3}$Astrophysics Research Centre, University of KwaZulu-Natal\\
$^{4}$South African Radio Astronomy Observatory, Liesbeek House, Settlers Way, Mowbray, Cape Town 7705, South Africa\\
$^{5}$Institute of Astronomy, National Tsing Hua University, Hsinchu 300044, Taiwan\\
$^{6}$School of Mathematics, Statistics, and Computer Science, University of KwaZulu-Natal}

\date{Accepted 2025 March 19. Received 2025 March 01; in original form 2024 August 09}

\pubyear{2024}

\begin{document}
\label{firstpage}
\pagerange{\pageref{firstpage}--\pageref{lastpage}}
\maketitle

\begin{abstract}
Multiwavelength studies of galaxy clusters are crucial for understanding the complex interconnection of the thermal and non-thermal constituents of these massive structures and uncovering the physical processes involved in their formation and evolution. Here, we report a multiwavelength assessment of the galaxy cluster A384, which was previously reported to host a radio halo with a 660 kpc size at MeerKAT 1.28 GHz. The halo is slightly offset from the cluster centre. Our uGMRT observation reveals that the halo extends up to 690 kpc at 407 MHz with a nonuniform spectral index $\alpha^{1284\ \mathrm{MHz}}_{407\ \mathrm{MHz}}$ distribution varying from flat (-0.5) to steep (-1.3) values. 
In addition, we use legacy GMRT 608 MHz, \textit{XMM-Newton} X-ray, and the Dark Energy Survey optical observations to obtain an extensive understanding of the dynamical nature of the galaxy cluster. The X-ray surface brightness concentration parameter ($C_{\rm{SB}} = 0.16$) and centroid shift ($w = 0.057$) reveal an ongoing dynamical disturbance in the cluster. This is also supported by the elongated 2-D optical galaxy density distribution map of the cluster. The obtained centre shift between optical and X-ray peaks and the asymmetry parameter from optical analysis further supports the dynamical disturbance in the cluster. The radio and X-ray surface brightness follows a sub-linear correlation. Our observations suggest that the cluster is currently in a merging state where particle re-acceleration in the turbulent ICM resulted in the radio halo emission.
\end{abstract}




\section{Introduction}
The theory of structure formation predicts that galaxy clusters grow via a combination of accretion of galaxies, galaxy groups, and smaller clusters. 
The merging processes make the clusters a rich centre of astrophysical interactions. One of the most intriguing aspects of galaxy cluster physics is the study of the cluster scale diffuse radio emission. Radio observations of clusters reveal the presence of steep spectrum ($\alpha<-1$, where $S_{\nu} \propto \nu^{\alpha}$), diffuse radio structures that are not necessarily associated with the cluster radio galaxies. The presence of these structures allows us to trace the history of cosmic ray acceleration, distribution of non-thermal relativistic electrons, and magnetic fields in the intracluster medium (ICM) \citep{Feretti_2012A&A, Brunetti_2014IJMPD}.
\begin{table}
\centering
\caption{Abell 384 Cluster properties including position, mass, luminosity and temperature as noted from \citetalias{Knowles_2021MNRAS}, \citealt{Piffaretti_2011A&A, Planck_Collaboration_2014AA, Bhargava_2020MNRAS}}
\label{tab:table1}
\begin{tabular}{c c} 
\hline
\hline
Parameter & Value\\
\hline
	RA$_{\mathrm{J2000}}$ & 02h 48m 11s \\
	DEC$_{\mathrm{J2000}}$ &  $-02\degree 16\arcmin\ 27\arcsec$\\
    Redshift ($z$) & 0.238\\
	Mass ($M_{\rm{SZ}}$) & $6.4 \pm 0.6 \times10^{14}\ \mathrm{M}_{\odot}$\\
	$L_{\rm{X}(0.1-2.4\ \mathrm{keV})}$ & 6.2$ \times 10^{44}\ \mathrm{h_{50}^{-2}\ erg\ s^{-1}}$\\
    $T_{\rm{X(0.3 - 7.9\ \mathrm{keV})}}$ & $7.7\pm 0.4$ keV\\
 \hline
\end{tabular}
\end{table}
\begin{figure*}
\includegraphics[width=\columnwidth]{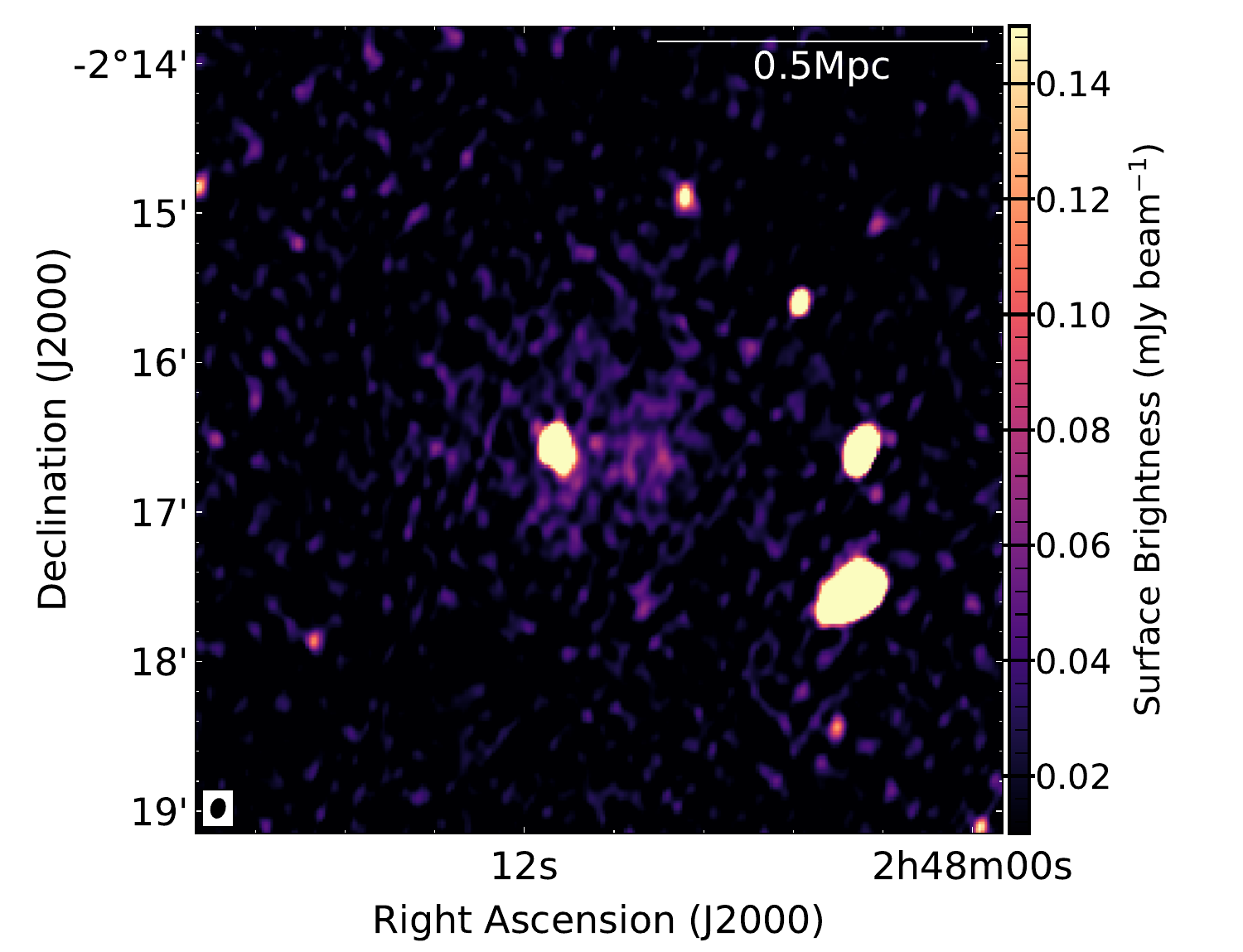}
\includegraphics[width=\columnwidth]{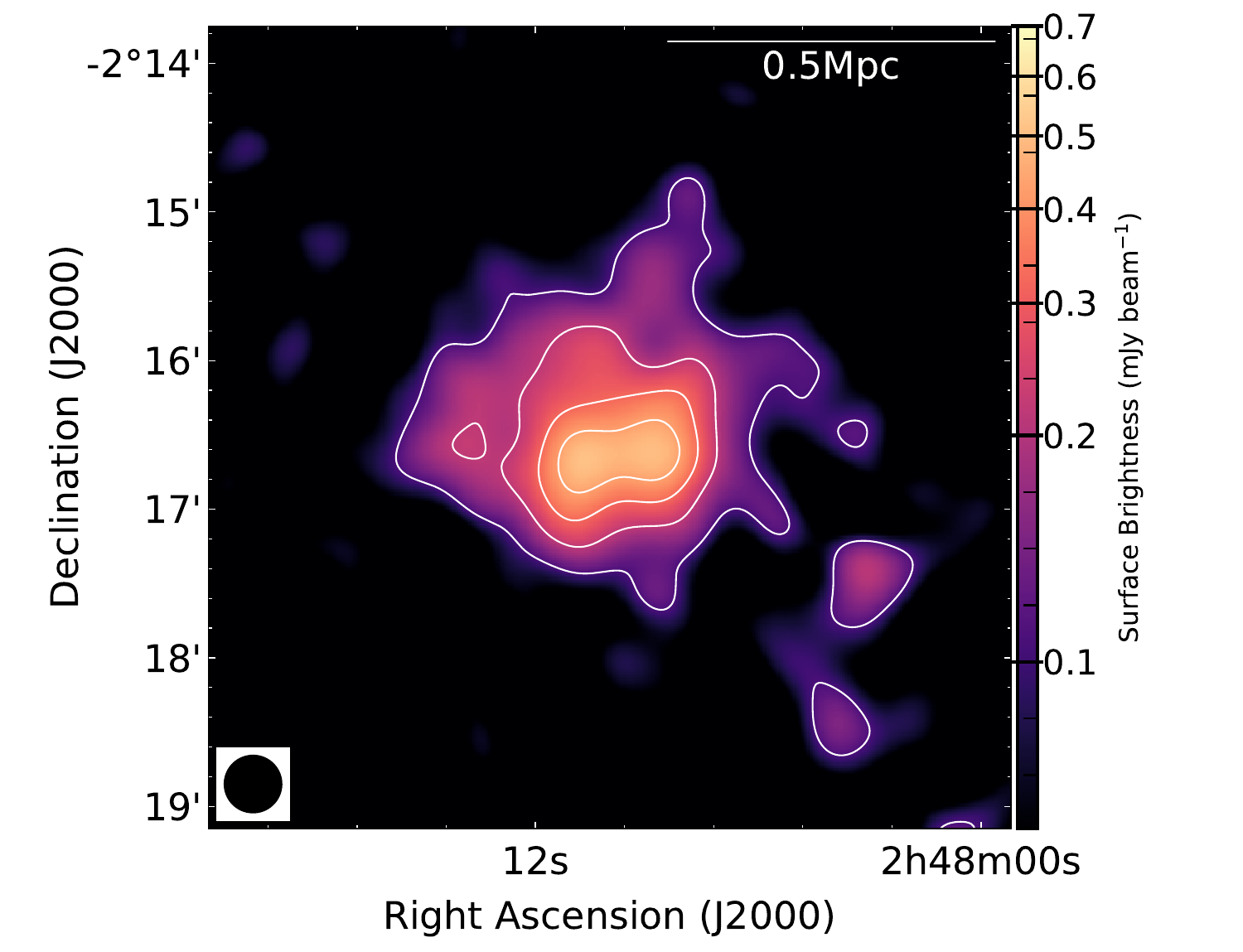}\\
\caption{Left: MeerKAT 1.28 GHz high-resolution image (Resolution: $7.7\arcsec \times 5.4\arcsec, -14.1\degree$) of A384. Right: Compact source subtracted low-resolution ($23\arcsec$) image of the radio halo at 1.28 GHz overlaid with low-resolution contours in white. The contours are placed at (3,6,9,12)$\sigma_{1.28~\rm{GHz}}$, where $\sigma_{1.28~\rm{GHz}} = 36 ~\mu$Jy/beam.}
\label{fig: J0248_radio_meerkat}
\end{figure*}
Radio Halos, mostly found in the centres of merging clusters, give evidence of a turbulent and disturbed ICM \citep{Cassano_2010ApJ}. They often trace the cluster X-ray emission. The theory of the hadronic model on the origin of radio halos suggests that the relativistic electrons are continuously injected in the ICM through the hadronic interaction of cosmic ray protons with ICM protons \citep{Dennison_1980ApJ, Schlickeiser_1987A&A, Blasi_1999APh, Brunetti_2017MNRAS}. However, the steep and non-uniform distribution of spectral index and non-detection of Gamma-ray emission from clusters support the theory of turbulent re-acceleration where electrons are re-energized to relativistic speeds by magneto-hydrodynamical turbulence generated during cluster mergers \citep{Petrosian_2001ApJ, Brunetti_2001MNRAS, Brunetti_2011MNRAS}. 
Another type of cluster-centric diffuse radio emission, the radio minihalos observed in cool core clusters, are an indicator of minor merger or gas sloshing inside the ICM \citep{Giacintucci_2011A&A, ZuHone_2013ApJ,Raja_2020ApJ}. Thus, the study of these diffuse radio emissions serves as a proxy for understanding clusters' dynamical history. However, the detection of radio halos in clusters with cool cores \citep{Bonafede_2014MNRAS, Sommer_2017MNRAS} highlights the limitations of relying solely on radio observations to infer a cluster's merger history. This necessitates multiwavelength observation, which helps in
understanding different physical interactions happening in the ICM and the interplay between the different cluster components \citep{Zenteno_2020MNRAS, Chatterjee_2022AJ,Ubertosi_2023A&A}.
Over the last few decades, clusters have been studied in all bands of the electromagnetic spectrum to probe the different aspects of the systems. For instance, we can map the distribution and velocity dispersions of member galaxies through optical observations, which is an excellent way to study the kinematics of the cluster galaxies \citep{Barrena2013MNRAS, Kim_2019ApJ}. Velocity dispersions also act as a tracer of the cluster's dynamical mass \citep{Ruel_2014ApJ, Farahi_2016MNRAS, Ferragamo_2021A&A, Chen_2022ApJ}. Optical density maps help search for signs of sub-structure, which provides hints about the merging history \citep{Girardi_2002ASSL, Foex_2019arXiv}. X-ray observations reveal the properties of the hot intracluster gas. Clusters' mass, temperature, density, history of heating and cooling etc., can be interpreted from X-ray observations of galaxy clusters \citep{Datta_2014ApJ, Yuan_2020MNRAS, Rahaman_2021MNRAS}. By combining these different observations, we can build a more complete picture of the physical properties of a cluster and its dynamical state.

\begin{table*}
	\centering
	\caption{The observation summary including the archival observation ID for each telescope, the central frequency and bandwidth, total on-source time of observation, high-resolution beam size and achieved RMS noise for off-source central region for MeerKAT and GMRT are listed below}
	\label{tab:Table2}
	\hspace{0.01in}
    \resizebox{\textwidth}{!}{
	\begin{tabular}{cccccccc} 
	    \hline
	    \hline
		Telescope & Observation & Central Frequency & Bandwidth & Time on Source & Restoring Beam & P.A. & RMS \\
		   & ID & (MHz) & ( MHz) & (min) & (high-resolution)& & ($\mu$ Jy/beam) \\
        \hline
        MeerKAT & SCI-20190418-KK-01 & 1284 & 855 & 24 & $7.7\arcsec  \times 5.5\arcsec $ & $-14.2$\degree & 14.5\\
		GMRT & 26\_021 & 608 & 32 & 190 & $10.7\arcsec  \times 6.7\arcsec $ & $-41.6$\degree & 40\\
        uGMRT & 40\_071 & 407 & 200 & 210 & $11.8\arcsec \times 10.9\arcsec $ & 87.8\degree & 60\\
	    \hline
	\end{tabular}
	}
\end{table*}

\begin{figure*}
\includegraphics[width=\columnwidth]{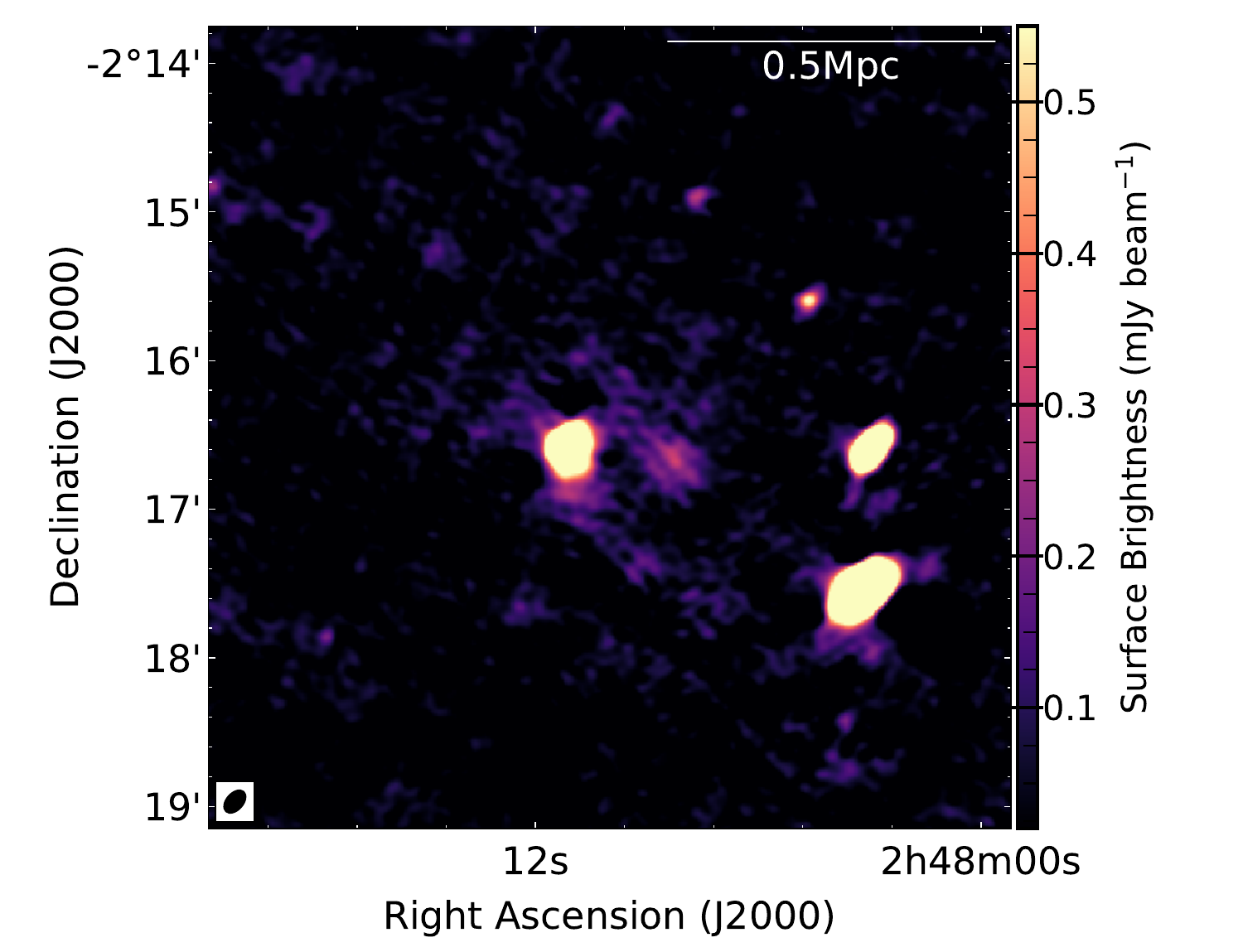}
\includegraphics[width=\columnwidth]{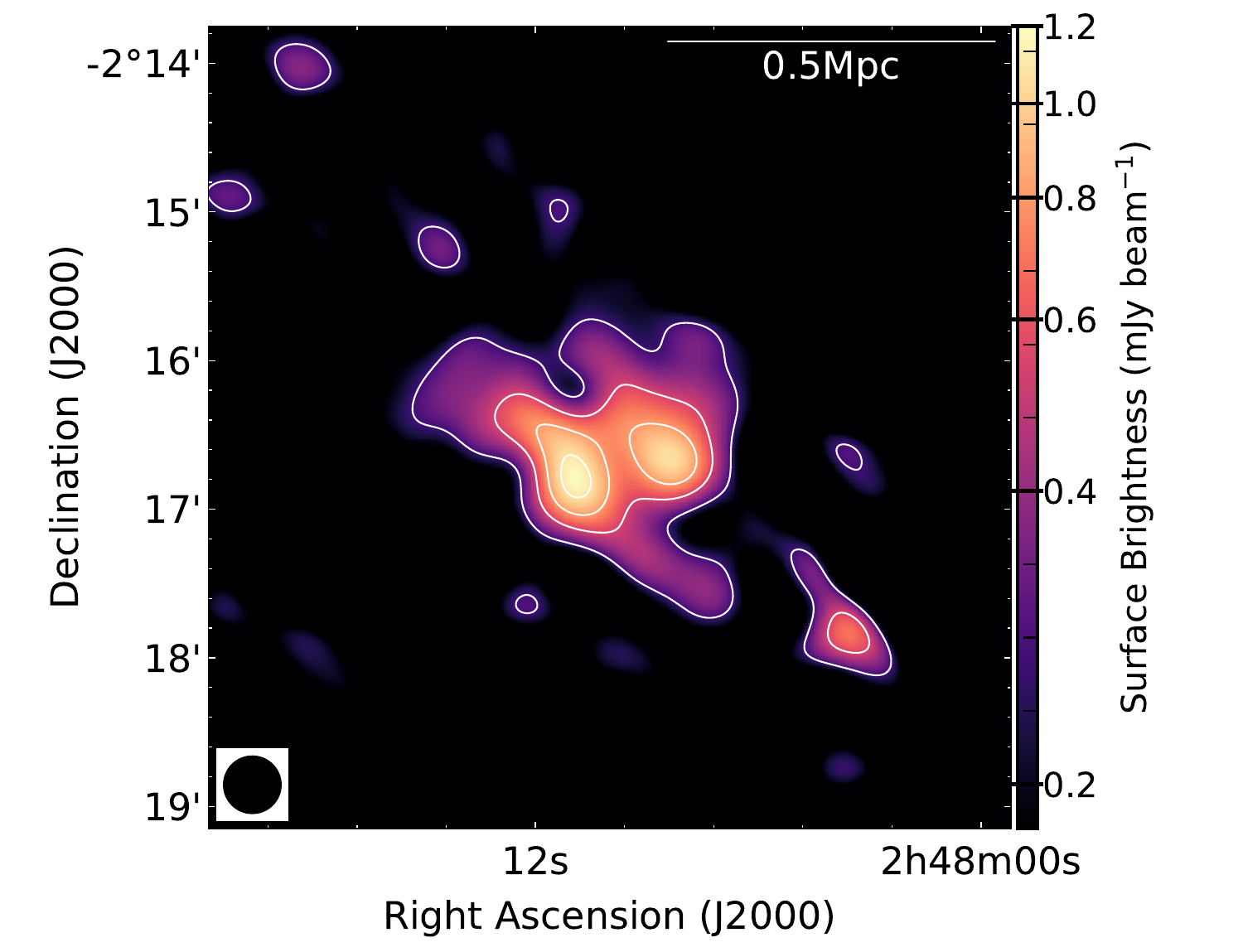}\\
\caption{Left: GMRT 608 MHz high-resolution image (Resolution: $10.6\arcsec \times 6.7\arcsec, -41.6\degree$) of A384. Right: Compact source subtracted low-resolution ($23\arcsec$) image of the radio halo at 608 MHz overlaid with low-resolution contours in white. The contours are placed at (3,6,9,12)$\sigma_{608~\rm{MHz}}$, where $\sigma_{608~\rm{MHz}} = 90 ~\mu$Jy/beam.}
\label{fig: J0248_radio_610}
\end{figure*}

\begin{figure*}
\includegraphics[width=\columnwidth, ]{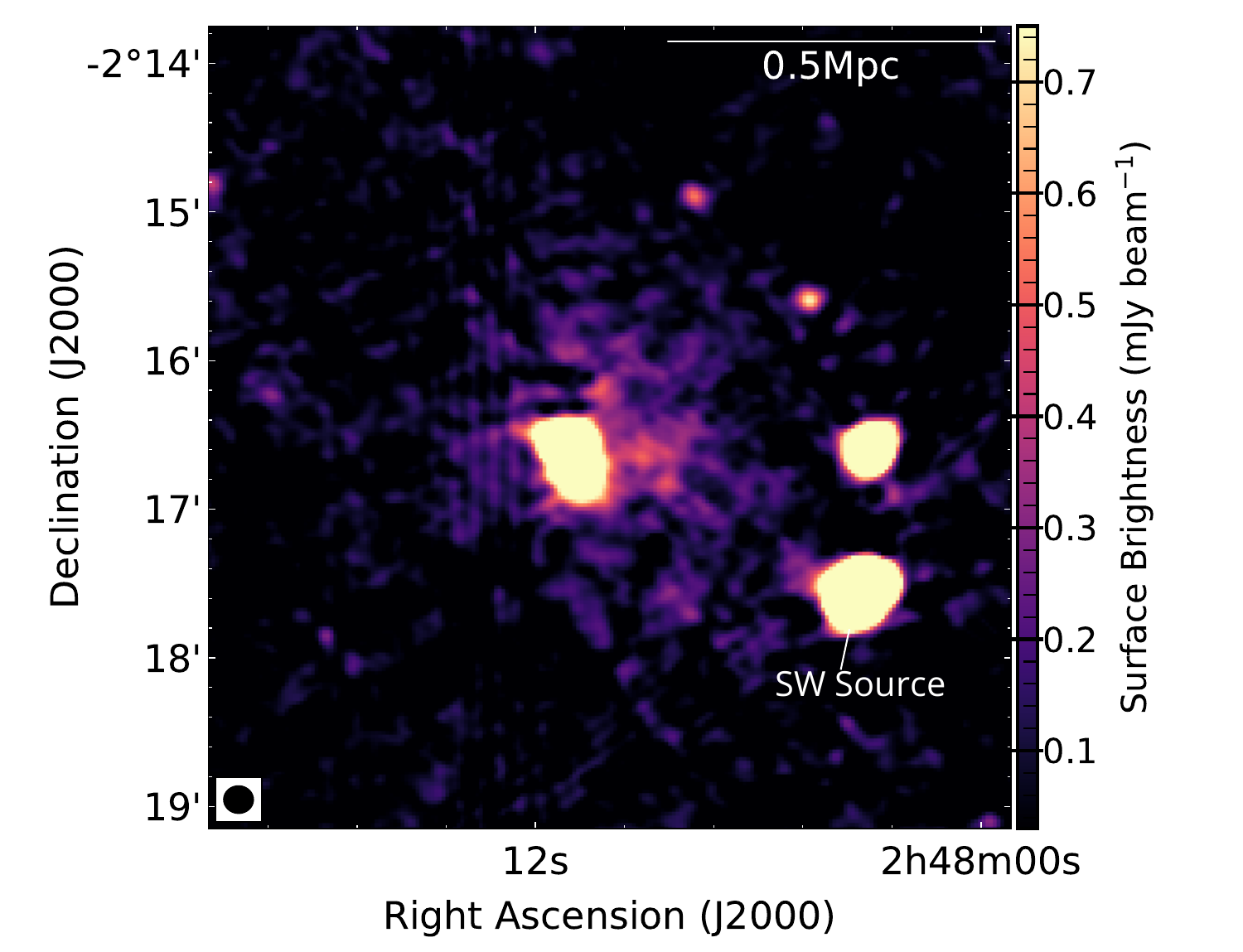}
\includegraphics[width=\columnwidth]{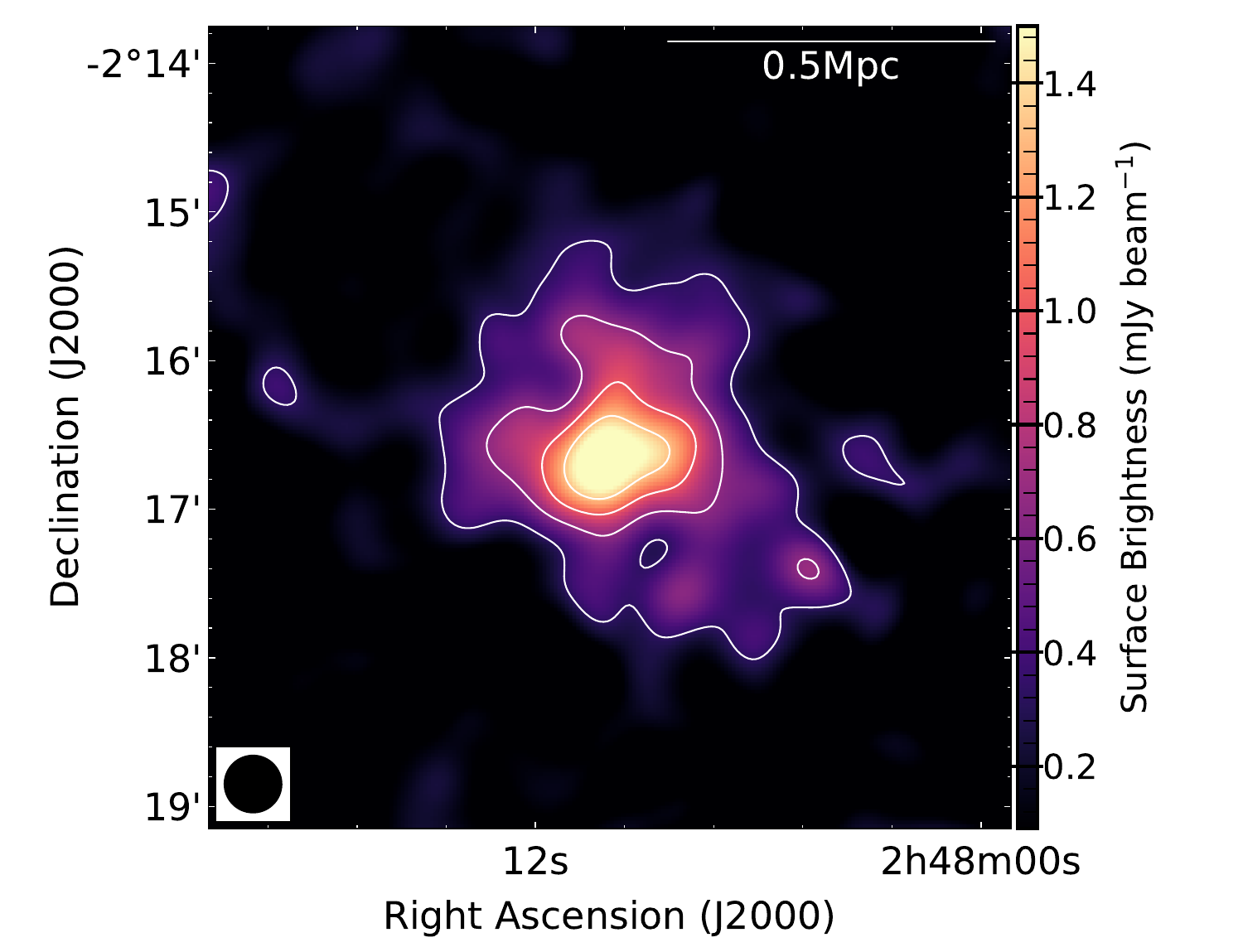}\\
\caption{Left: uGMRT 407 MHz high-resolution image (Resolution: $11.8\arcsec \times 10.9\arcsec, 87.8\degree$) of A384. Right: Compact source subtracted low-resolution ($23\arcsec$) image of the radio halo at 407 MHz overlaid with low-resolution contours in white. The contours are placed at (3,6,9,12)$\sigma_{407~\rm{MHz}}$, where $\sigma_{407~\rm{MHz}} = 110 ~\mu$Jy/beam.}
\label{fig: J0248_radio_400}
\end{figure*}

Abell 384 or ACT-CL J0248.1-0216 (hereafter A384) is a massive cluster located at a redshift of 0.238. The cluster was reported to have a  luminosity $L_{\rm{X~(0.1-2.4 keV)}} = 6.2 \times 10^{44}~\mathrm{erg.s^{-1}}$ \citep{Piffaretti_2011A&A} and temperature $T_{\rm{X~(0.3 - 7.9~keV)}} = 7.7\pm 0.4$ keV \citep{Bhargava_2020MNRAS} from X-ray observations, and a mass $M_{\mathrm{SZ}} = 6.4 \pm 0.6 \times 10^{14}~M_\odot$ form Sunyaev–Zeldovich effect (SZ) observations \citep{Planck_Collaboration_2014AA}. The cluster properties are summarised in Table \ref{tab:table1}. 
The cluster was detected hosting a 660 kpc radio halo in the MeerKAT Exploration of Relics, Giant Halos, and Extragalactic Radio Sources (MERGHERS) pilot survey at MeerKAT L band \citep[][hereafter \citetalias{Knowles_2021MNRAS}]{Knowles_2021MNRAS}. In this paper, we discuss the spectral behaviour and dynamical nature of the cluster using a combination of radio, X-ray and optical studies. We present the observations and data reduction procedures in Section ~\ref{radio_obs}. 
In Section ~\ref{radio_results}
we present the cluster diffuse emission properties obtained from radio observation. We present the results from X-ray and optical observations in Section ~\ref{X-ray results} and in Section ~\ref{optical_results}, respectively. We present the discussion regarding the dynamical state of the cluster in Section \ref{discussion}, and a summary in Section ~\ref{summary}.\\
Assuming a $\Lambda$CDM cosmology with $\Omega_m$ = 0.3, $\Omega_{\Lambda}$ = 0.7 and $H_0$ = 70\,km\,s$^{-1}$\,Mpc$^{-1}$ at the redshift of the Abell 384 ($z = 0.238$), 1\arcsec = 3.77\, kpc.


\section{Observations and Data Reduction}\label{radio_obs}
In this section, we discuss the observations and multiwavelength data analysis of A384 using radio, X-ray and optical observations. We use MeerKAT L band (1.28 GHz), upgraded Giant Metrewave Radio Telescope (uGMRT) band-3 (407 MHz), and archival GMRT 608 MHz radio observations to study the radio spectral behaviour. In addition, we use XMM Newton X-ray and Dark Energy Survey (DES) optical observations to study the cluster dynamics in detail.

\subsection{Radio Observations}
Radio observations of the cluster using MeerKAT and Giant Metrewave Radio Telescope (GMRT) and the respective data reduction procedures are presented in this section. The summary of the observations is presented in Table \ref{tab:Table2}.

\subsubsection{MeerKAT 1.28 GHz}\label{meerkat}
A384 was observed with MeerKAT L band (central frequency 1.28 GHz) with a 24 minutes on-source time as a part of the MERGHERS Pilot survey (\citetalias{Knowles_2021MNRAS}) where a 660 kpc radio halo was discovered. The data reduction for the observation was carried out using the semi-automated \textsc{OXKAT}\footnote{\url{https://github.com/IanHeywood/oxkat}} pipeline \citep{Heywood_2020ascl}, which performs the initial direction-independent calibration. Moreover, to account for the direction-dependent effect, \textsc{KillMS} \citep{Tasse_2014ARXIV} and \textsc{DDFACET} \citep{Tasse_2018A&A} were used. The data was affected by RFI at small baselines, leading to the flagging of baselines below $180\lambda$. The high-resolution image (at 1.28 GHz) used here was made in \textsc{WSClean} \citep{offringa_2014MNRAS, Offringa_2017MNRAS}, using an uv-range of $180\lambda - 41k\lambda$ with Briggs robust $-0.5$ \citep{Briggs_1995AAS} and was restored with a beam size of $7.7\arcsec, 5.4\arcsec, -14.1\degree$ (Figure ~\ref{fig: J0248_radio_meerkat}, Left). The detailed analysis of the MeerKAT observations and results are presented in \citetalias{Knowles_2021MNRAS}. Here, we have used the previously calibrated high-resolution image of A384 presented in \citetalias{Knowles_2021MNRAS} for further analysis.

We subtracted the compact sources not associated with the radio halo emission using the image plane filtering technique \citep{Rudnick2002PASP}. To proceed with the technique, we used the high-resolution image of the cluster and processed it further using the \textsc{radio source analysis code}\footnote{\url{https://github.com/twillis449/radio_source_analysis}}. A brief procedure of how the code works is summarised as follows. At first, The compact sources are identified from the image by the \textsc{breizorro masking program}\footnote{\url{https://github.com/ratt-ru/breizorro}}, where pixels brighter than 5 times the image noise are identified as sources for masking. The sources falling within 1$-$3 times the beam size were filtered with a disk of size 3. This disk removes flux from all the unresolved sources falling within the specified size successfully ($>$90\% flux from the central radio galaxy). The filtered image is then corrected for the zero level. The final output was then convolved to lower resolutions of 23$\arcsec$ to obtain the halo emission (Figure \ref{fig: J0248_radio_meerkat}, Right).
\begin{table}
\centering
\caption{For each observing frequency, the obtained low-resolution RMS noise, the LLS and flux densities obtained within $3\sigma$ region for radio halo are mentioned.}
\label{Tab: diffusetable}
\resizebox{\columnwidth}{!}{
\begin{tabular}{cccccc}
\hline
Frequency & Beam & RMS & LLS & Flux Density \\
(MHz) & low-resolution&($\mu$Jy/beam) & (kpc) & (mJy)\\
\hline
1284 & $23 \arcsec$ & 36 & 660 & $5.72 \pm 0.32$ \\
608 & $23 \arcsec$ & 90 & 460 & $8.98\pm0.79$ \\
407 & $23 \arcsec$ & 110 & 690 & $17.21\pm1.33$ \\ \hline
\end{tabular}}
\end{table}
\subsubsection{Archival GMRT 608 MHz}
We made use of the available archival 608 MHz observation (Project: 26\_021) of A384. The observation was performed for 190 minutes on-source time using legacy GMRT software backend (GSB) 608 MHz band where the 32 MHz bandwidth is split over 216 channels. The Source Peeling and Atmospheric Modelling (\textsc{SPAM}) pipeline \citep{Intema_2009, Intema_2017} was used for data processing and imaging with a Briggs robust 0. \textsc{SPAM} is a semi-automated pipeline that takes care of RFI removal and standard interferometric calibration.  
\textsc{SPAM} also takes care of direction-dependent effects and produces a primary beam-corrected image along with the calibrated visibility file at the end. We used the calibrated \textsc{SPAM} output in FITS format and converted it to CASA measurement sets for further analysis. The imaging was performed using \textsc{WSClean}.

We created high-resolution images of the clusters with the same UV range of $180\lambda - 41k\lambda$ as used for MeerKAT images by \citetalias{Knowles_2021MNRAS} (Figure ~\ref{fig: J0248_radio_610}, Left). The resolution and final achieved RMS for each cluster are mentioned in Table \ref{tab:Table2}. The central UV coverage of MeerKAT is much denser than GMRT. Thus, to make the images from both arrays comparable, the 608 MHz GMRT image was created using robust $+0.5$ (Figure ~\ref{fig: J0248_radio_610}, Left). The compact sources thereafter were subtracted from the image as described in Section~\ref{meerkat}. The convolved residual image having the halo emission is presented in Figure \ref{fig: J0248_radio_610} (Right).

\subsubsection{uGMRT 407 MHz}

As a low-frequency follow-up, we observed A384 with uGMRT band-3 as a part of Project 40\_071 with a total of 3.5 hours of on-source time. The observation was taken in both the GSB and the GMRT wideband backend (GWB) mode. The GWB observation was performed in full polarisation mode, where the 200 MHz bandwidth was split into 4096 channels. Wideband uGMRT observations are frequently impacted by undesired radio frequency interference (RFI). Real-time RFI filtering \citep{Buch2022JAI} was applied during the observation to mitigate those effects.

The CAsa Pipeline-cum-Toolkit for Upgraded GMRT data REduction (CAPTURE\footnote{\url{https://github.com/ruta-k/uGMRT-pipeline}}; \citealt{Kale_2021ExA_capture}) pipeline was used for data reduction. CAPTURE follows the standard procedure of data reduction, where initially, the bad channels and scans were flagged. Subsequently, the delay, bandpass, and initial gain calibrations were performed. The gain calibration solutions from phase calibrators were then applied to the target source. The target data was averaged and flagged for any remaining corrupted data. Following this, two rounds of self-calibration were executed to correct for residual phase errors. Finally, a multi-scale multi-frequency synthesis (MS-MFS) map for the cluster was created using the joined channel deconvolution mode with \textsc{WSClean}, using an uvrange of $180\lambda - 41k\lambda$ and robust 0.5. Thereafter, the primary beam correction was done using \textsc{CASA} task \texttt{wbpbgmrt}. 

The final image achieved is presented in Figure ~\ref{fig: J0248_radio_400} (Left). This image was further used for compact source subtraction following the method described in section Section~\ref{meerkat}. The radio halo emission observed at 407 MHz is shown in Figure~\ref{fig: J0248_radio_400} (Right).


\subsection{X-ray Observation}\label{xray_obs}
\begin{figure}
\includegraphics[width=\columnwidth]{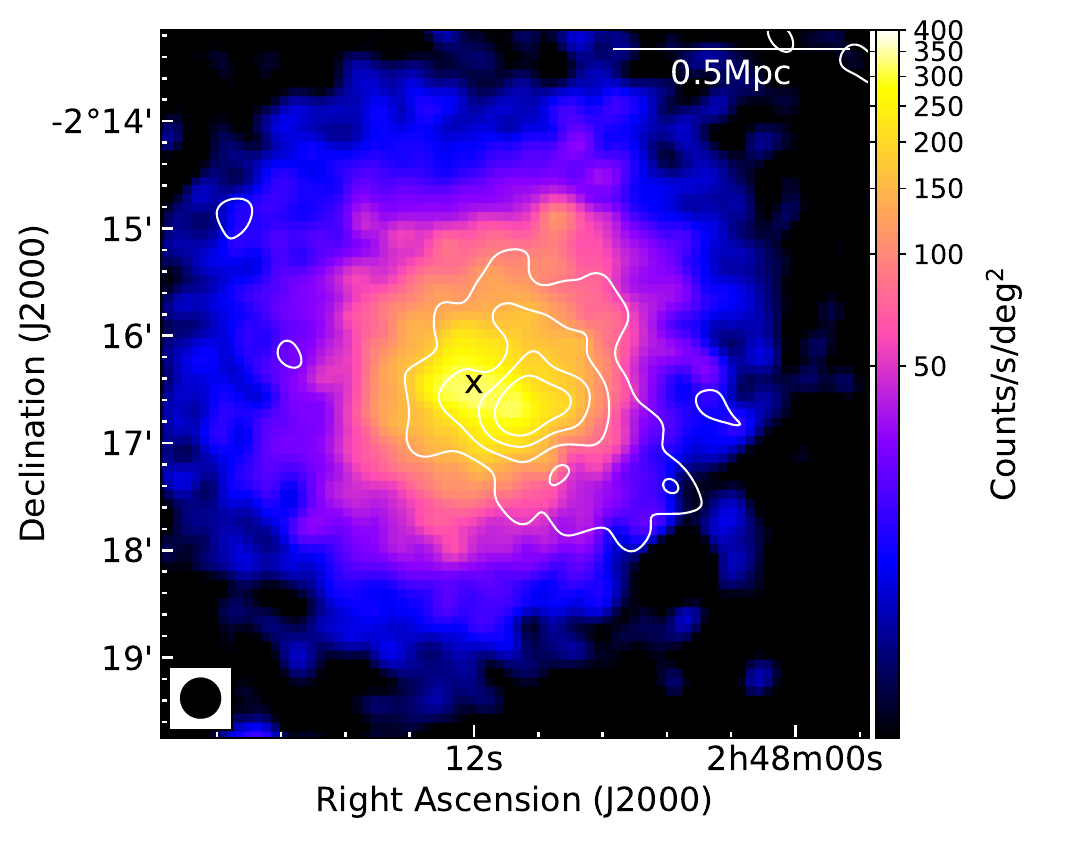}\\
\caption{A384 XMM Newton X-ray SB map overlaid on uGMRT 407 MHz low-resolution image contours in white. The contours placed from (3,6,9,12)$\sigma$, where $\sigma = 110\ \mu$Jy/beam. The X-ray peak is marked with black `x'.}
\label{fig: A384_xray}
\end{figure}

\begin{figure*}
\includegraphics[width=\columnwidth]{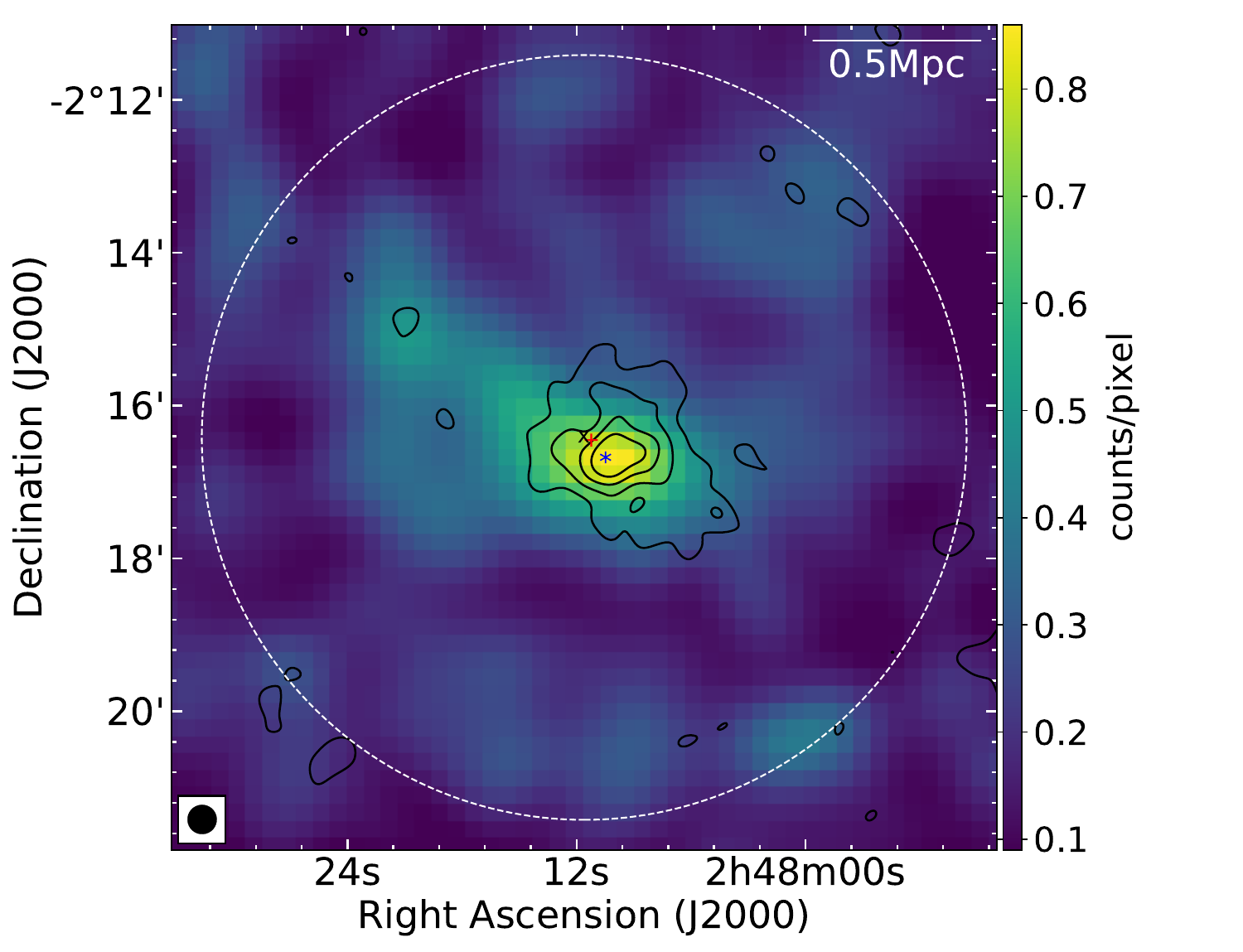}
\includegraphics[width=\columnwidth]{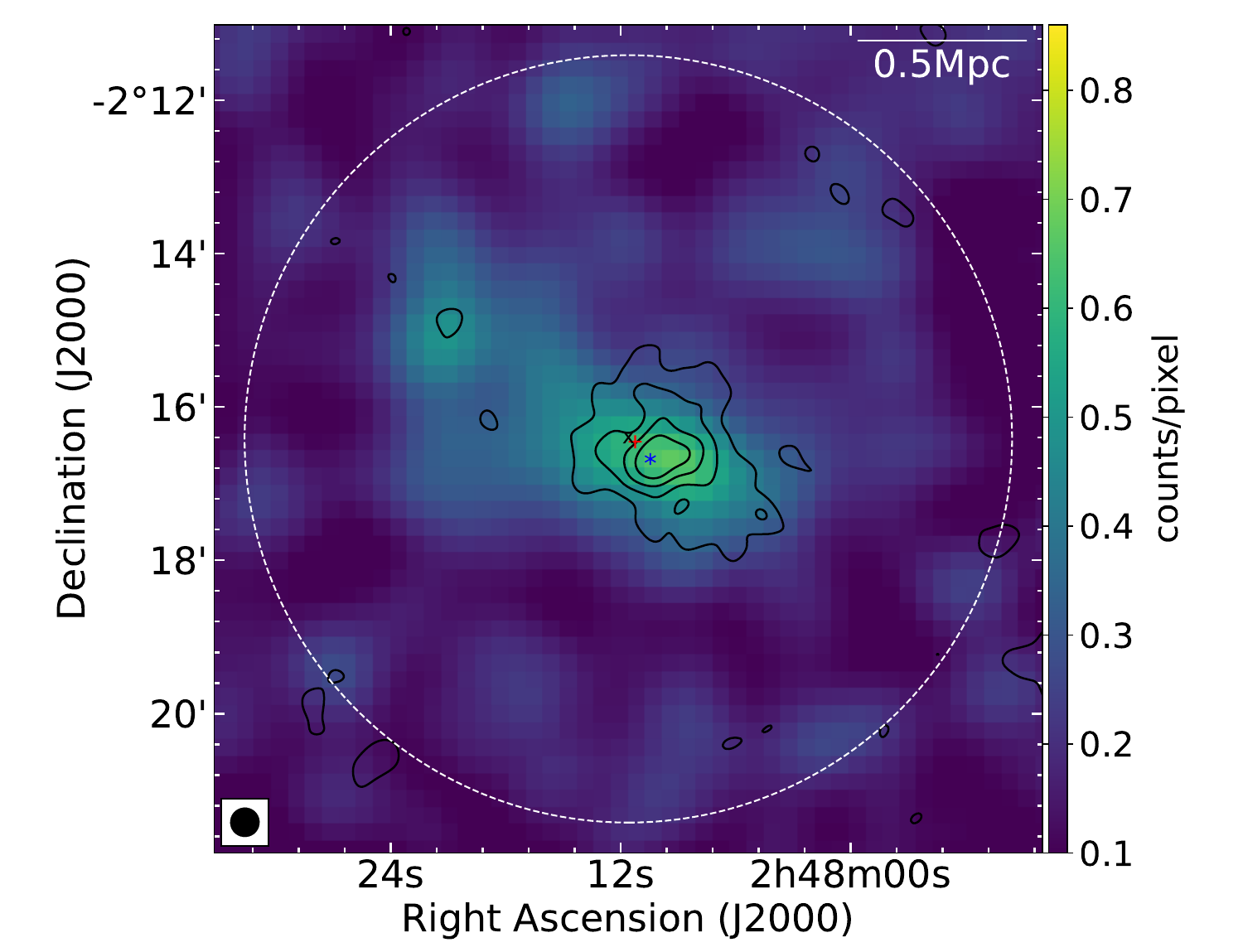}\\
\caption{Left: 2.5 $\times$ 2.5 Mpc projected density map of A384 obtained from \textit{zField} using photometry from the DES DR1 database. Right: Corresponding error map. Each pixel in the projected maps is 0.05 Mpc $\times$ 0.05 Mpc in size. uGMRT 407 MHz low-resolution contours are overlaid in black and the R$_{500}$ region of the cluster is marked with a white dashed circle. The SZ peak of the cluster from \citet{Hilton_2018ApJS} and the X-ray peak are marked with red `+' and black `x'. The peak of the optical density map is marked with blue `*'}. 

\label{fig: A384_opticaldensity}
\end{figure*}

We have used the archival 40.6 ks XMM-Newton space telescope observation (obs ID 0721890401) of A384 to study the X-ray morphology of the cluster. The XMM Newton Extended Source Analysis Software (\textsc{ESAS}) package available in the Science Analysis Software (\textsc{SAS}) v20\footnote{\url{https://www.cosmos.esa.int/web/xmm-newton/sas}} was used for the analysis. The data analysis was done in the following manner. First, the calibration files and observation data were generated using the \textit{cifbuild} and \textit{odfingest} tasks. The event files for PN and MOS observations were generated using \textit{epchain} and \textit{emchain} tasks. CCD 3 and 6 for MOS1 were excluded for being in the anomalous state. CCD 4 for MOS 2 was excluded for having bad data below 1 keV. \textit{mos-filter} and \textit{pn-filter} were used to remove soft proton flares. Point sources were masked and excised using the \textit{cheese} tasks. Next, pn-spectra (mos-spectra) and pn-back (mos-back) were used to process filtered event files to create the source and model background spectra as well the EPIC images in the energy range of 0.4-2 keV. The spectral data was grouped using the FTOOL
\textsc{grppha} included in \textsc{Heasoft}. The X-ray background below 2 keV is dominated by soft emission from within the Milky Way. We downloaded the relevant XRBG files and RASS spectrum and appropriate response matrices from the HEASARC X-ray background tool\footnote{\url{https://heasarc.gsfc.nasa.gov/cgi-bin/Tools/xraybg/xraybg.pl}}. \textsc{XSPEX} v12 was used for spectral fitting. Thereafter \textit{sp\_partial} was used to scale the
\textsc{XSPEX} normalizations from the cheesed data to the full field of view. The final exposure corrected background subtracted 0.4-2 keV image was created using ESAS task \textit{comb} and \textit{adapt} (Figure \ref{fig: A384_xray}).

\subsection{Optical Observation}\label{optical_obs}

For optical analysis, the available DES Data Release 1 data has been used. We used the \textsc{zCluster} code \citep{Hilton_2018ApJS} for estimating the photometric redshift of galaxy clusters and creating 2-D projected density maps. The script uses a template fitting method to determine the redshift probability distributions of individual galaxies \citep{Benitez2000ApJ}. The probability distribution for each galaxy was integrated around \textit{z} = $0.238 \pm 0.2$ to calculate the weighting for the density map with 0.238 being the cluster redshift (\citetalias{Knowles_2021MNRAS}). Thereafter, the \textit{zfield} command was used to create a 2-D projected density map where the cluster's RA, DEC, and redshift information, along with the probability distribution catalogue, are utilised. Following \citet{Wen_Han2013MNRAS}, each galaxy was positioned on the 2-D map according to its sky position, where the celestial coordinates were mapped in Cartesian coordinates with the help of tangent plane projection. The resulting map is convolved with a Gaussian kernel to smooth it, and a density uncertainty map is created using Monte Carlo simulations where 68\% of the map is taken as an uncertainty map after 1000 iterations (Figure ~\ref{fig: A384_opticaldensity}).

\section{Results from Radio Observations}\label{radio_results}

\begin{figure}
\includegraphics[width=0.4\textwidth]{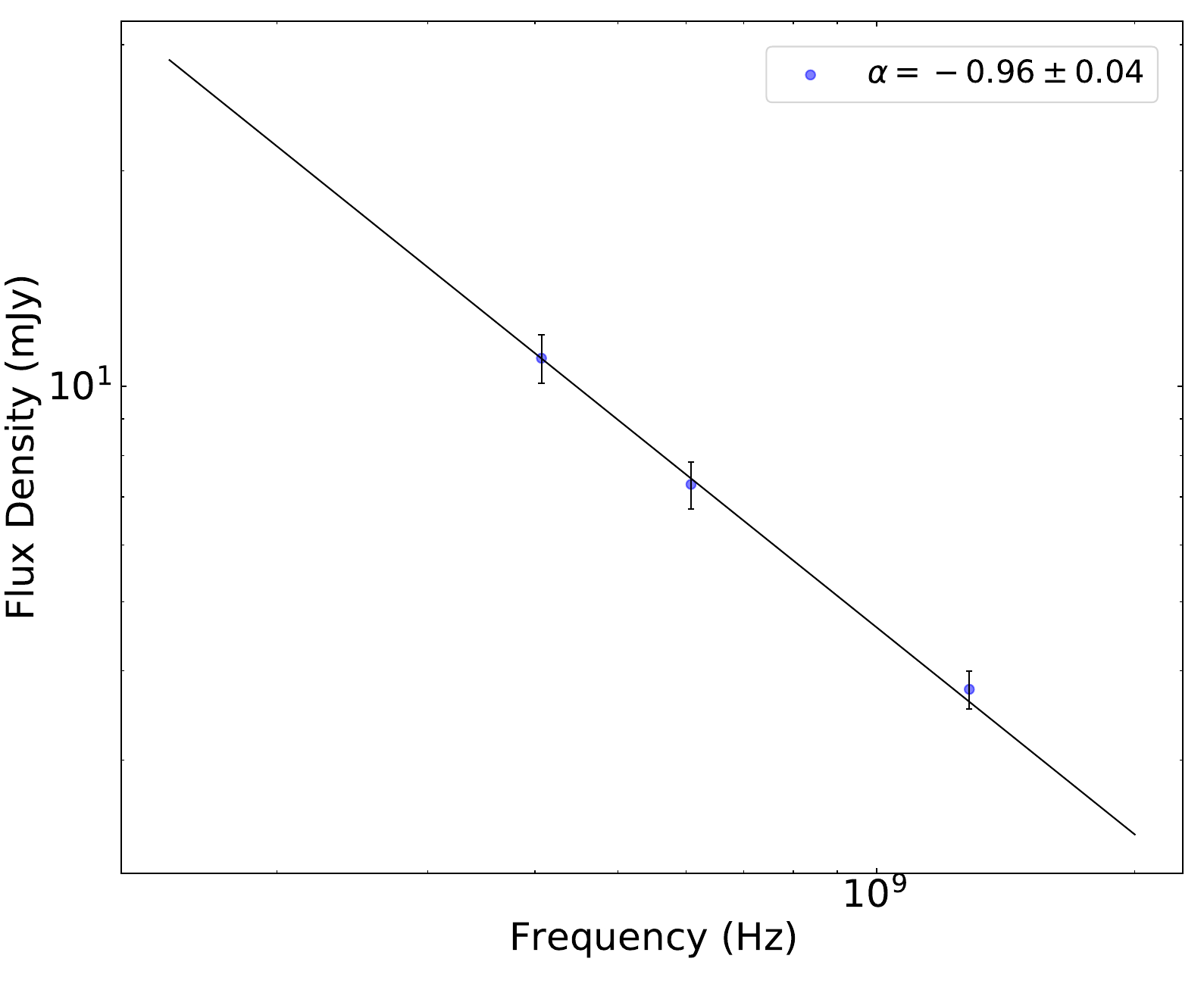}
\caption{The integrated flux densities falling with 3$\sigma$ region of 407 MHz, 608 MHz and 1.28 GHz of the radio halo plotted with frequency. The solid line marks the best-fit line and shows that the radio halo spectrum follows a power law between 407 MHz to 1.28 GHz with $\alpha = -0.96\pm0.4$}
\label{fig: A384_powlaw}
\end{figure}

The compact source subtracted low-resolution images of the clusters presented reveal the radio halo structure at each frequency. The low-resolution image RMS noise and the radio halo properties at each frequency are tabulated in Table \ref{Tab: diffusetable}. As mentioned earlier, the largest linear size (LLS) of the cluster detected by MeerKAT 1.28 GHz observation is 660 kpc. The halo has a much smaller extension of 460 kpc at 608 MHz, which is due to the insufficient sensitivity of the legacy GMRT observation. At 407 MHz, the halo extends up to an LLS of 690 kpc, which is slightly larger than the largest size detected at 1.28 GHz. The halo shows an elongation in the northeast-southwest direction at 407 MHz, which was not observed previously at 1.28 GHz. Furthermore, the halo does not follow the X-ray emission in the cluster and rather shows a slight offset (Figure \ref{fig: A384_xray}). The compact radio source at the southwest of the halo, marked as the SW source (Figure ~\ref{fig: J0248_radio_400}, Left), overlaps with an NVSS source with a possible optical host SDSS J024803.26-021728.9 at redshift 0.329 and thus is probably not associated with the halo emission.
\subsection{Flux Density and Integrated Spectrum}
The integrated flux density of the radio halo in the cluster was obtained by selecting the regions within the $3\sigma_\mathrm{rms}$ contour at each of the frequencies. At 1.28 GHz and 608 MHz, we obtained a flux density of $5.72 \pm 0.32$ mJy and $8.98\pm 0.79$ mJy, respectively.  At 407 MHz, the integrated flux density obtained is $17.21 \pm 1.33$ mJy. The uncertainty in flux density was estimated following \citep{vanweeren_2021A&A} using the equation
\begin{equation} \label{eq1}
\Delta S=\sqrt{(\sigma_\mathrm{cal}\ S)^2+(\sigma_\mathrm{rms} \sqrt{N_\mathrm{beam}})^2+\sum_{i} N_{\text{beams, i}} \sigma_{\text{rms}}^2} 
\end{equation}
where the flux calibration uncertainty ($\sigma_\mathrm{cal}$) of $5\%$ at $1.28$ GHz (\citetalias{Knowles_2021MNRAS}), $6\%$ at $608$ MHz \citep{chandra_2004ApJ} and $7\%$ at $407$ MHz \citep{Chatterjee_2024MNRAS} was considered. ${N_\mathrm{beam}}$ is the number of beams within the $3\sigma_\mathrm{rms}$ contour. The third term represents the uncertainty due to compact source subtractions where $\sum_{\text{i}} N_{\text{beams, i}}$ sums for the regions encompassing compact sources.
\begin{figure}
\includegraphics[width=0.9\columnwidth]{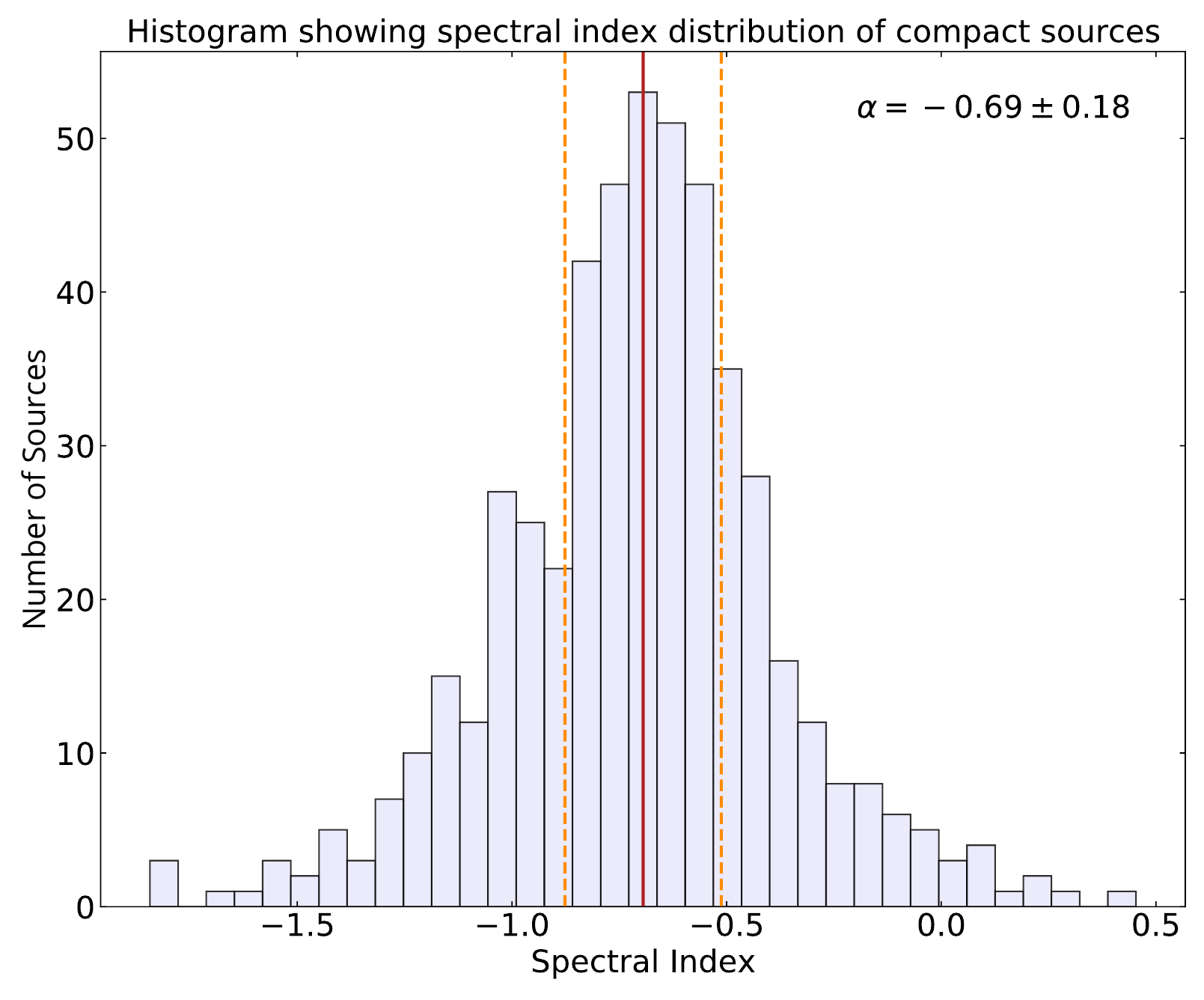}\\
\caption{The spectral index distribution of compact sources between 407 MHz and 1.28 GHz has a median spectral index $\alpha_{median} = -0.69$ with a median absolute deviation (MAD) 0.18.}
\label{fig: A384_407_1284_hist}
\end{figure}
\begin{figure*}
\includegraphics[width=0.48\textwidth]{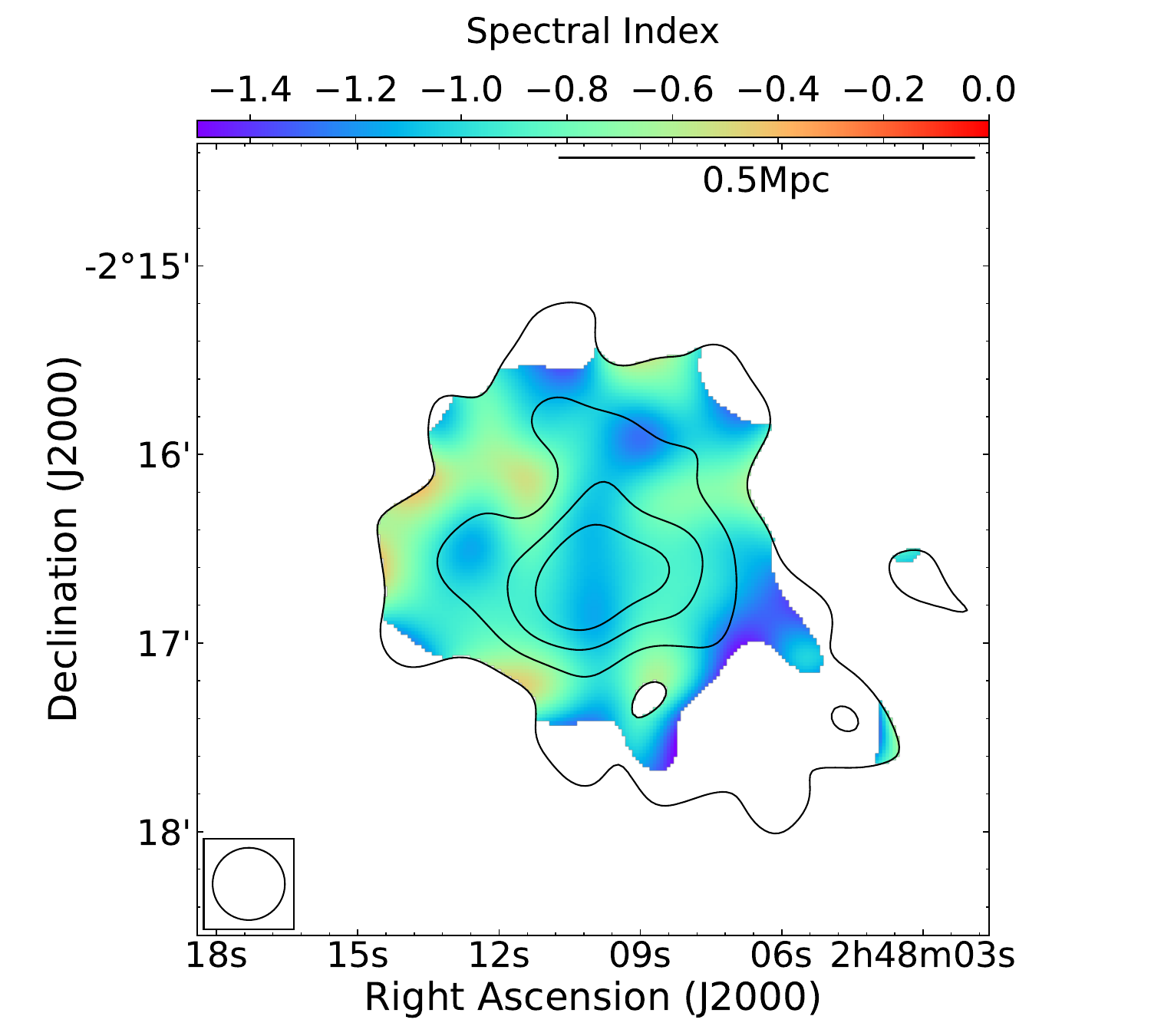}
\includegraphics[width=0.48\textwidth]{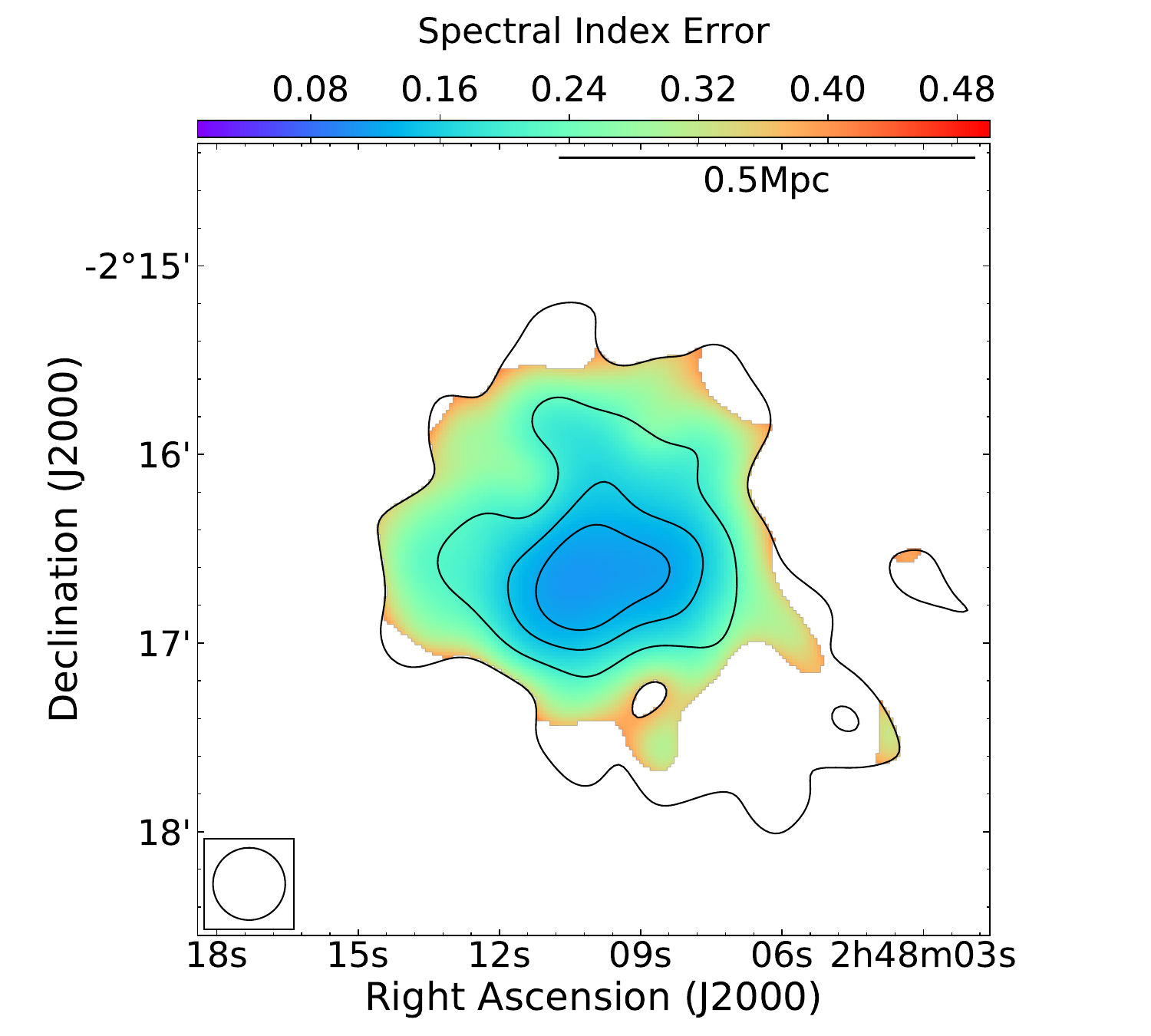}\\
\caption{A384 spectral index map between 407 MHz and 1.28 GHz overlaid with uGMRT 407 MHz low-resolution image contours. The contours placed from (3,6,9,12)$\sigma$, where $\sigma = 110~\mu$Jy/beam }
\label{fig: A384_spix}
\end{figure*}
We have measured the integrated spectral index of the radio halo between the three frequencies. The halo morphology differs at different frequencies. Therefore, while measuring the spectral index, we chose the common region with diffuse emission present above 3$\sigma$ significance at all three frequencies. The halo emission was found to follow a power law from 407 MHz to 1284 MHz with a spectral index $\alpha^{1284}_{407} = -0.96 \pm 0.04$ (Figure \ref{fig: A384_powlaw}).

The obtained spectral index value is slightly flatter compared to most observed radio halos, which typically exhibit steeper spectral indices ($\alpha< -1.1$; see \citealt{Feretti_2012A&A, vanweeren_2019SSRv} for a review). Moreover, the modern upgraded observations have led to the discovery of an increasing number of ultra-steep spectrum ($\alpha< -1.6$) radio halos \citep{Pasini_2024A&A}. To verify if the flatness is coming from any flux measurement error due to calibration, we checked the spectral index of the radio galaxy present in the primary beam corrected field of view in the 407 MHz and 1.28 GHz images. We measured the spectral index distribution of the compact sources in the field and found a median spectral index value of $-0.69$ for the compact sources with a median absolute deviation (MAD) of $0.18$ (Figure \ref{fig: A384_407_1284_hist}). This is consistent with the spectral index of radio galaxies observed in deep field studies \citep{Sinha_2023MNRAS}. 

Considering the spectral index of $-0.96$, we obtained the radio halo power at 1.4 GHz using equation \ref{eq3}.

\begin{equation} \label{eq3}
P_{1.4 \mathrm{\ GHz}}= \frac{4\pi D_L^2(z)}{(1+z)^{(\alpha+1)}}\times \Big(\frac{1400}{407}\Big)^{-0.96} \times S_{407\mathrm{\ MHz}}
\end{equation}

Where $D_L(z)$ is luminosity distance at the redshift $z$. \citet{Cuciti_2021A&A}, with a sample of 75 galaxy clusters having Planck SZ masses (M$_{500}>6 \times 10^{14}~\mathrm{M}_{\odot}$) updated radio power-vs-mass (P$_{1.4 ~\rm{GHz}}-M_{500}$) correlation of radio halos.
The estimated radio power of A384 $P_{1.4~\mathrm {GHz}} = 8.56\pm 0.51 \times 10^{23}~\mathrm{WHz^{-1}}$ is in agreement with the observed scaling relation ($P_{1.4 ~\rm{GHz}}$-$M_{500}$) by \citet{Cuciti_2021A&A}. Moreover, it also obeys the radio power- X-ray luminosity ($P_{1.4 ~\mathrm{GHz}}-L_{\rm{X}}$) scaling relations of the known radio halos \citep{cassano_2013ApJ,raja_2021MNRAS}. We further extrapolated the radio power to 150 MHz and obtained $P_{150~\mathrm {MHz}} = 7.55\pm 0.6 \times 10^{24}~\mathrm{WHz^{-1}}$. This obtained radio power is in alignment with the $P_{150~\mathrm{MHz}}-M_{500}$ correlation observed by \citet{Cuciti_2023A&A}.

\subsection{Spectral Index Map at 407 MHz and 1.28 GHz}
The spectral index variation and distribution of cluster radio halos serve several pieces of information regarding the origin of the halos and the particle acceleration history. We created a spectral index map of A384 between 407 MHz \& 1.28 GHz. As the halo emission is contained in a much smaller region at 608 MHz, we exclude this image from our further spectral analysis. 
To understand the spectral behaviour of the radio-emitting relativistic particles and compare the variation of spectral index across the halo, we created the spectral index map selecting regions falling within 3$\sigma$ contours for the images and masked the rest. The different images were made using different cell sizes; we used \textsc{CASA} task \textit{imregrid} on the 407 MHz map to convert the pixel grids and equalize them to that of the 1.28 GHz map. Thereafter, the spectral index (Figure \ref{fig: A384_spix}, Left) and spectral index error (Figure \ref{fig: A384_spix}, Right) maps were created using the \textit{immath} tool in \textsc{CASA}. The error map was calculated for each pixel using the following equation \\
\begin{equation} \label{eq2_s}
\Delta \alpha=\frac{1}{\rm{log}(\nu_1/\nu_2)}\times \sqrt{\frac{(\Delta S_1)^2}{S_1^2}+\frac{(\Delta S_2)^2}{S_2^2}}
\end{equation}

Figure \ref{fig: A384_spix} (Left) shows the spectral index distribution of the radio halo between 407 MHz \& 1.28 GHz at a resolution of $23\arcsec$. The overall spectral index distribution shows a clumpy structure with the spectral index values varying from $-0.5$ to $-1.3$, indicating the turbulent re-acceleration of particles in the disturbed ICM. Notably, no particular radial steepening was observed for this halo. We present further verification of the spectral index distribution in Appendix~\ref{Appendix}. The halo is more extended at 407 MHz in the southwest direction, which was not detected by MeerKAT. The particle acceleration efficiency in merger-generated turbulence is low, particularly in the cluster outskirts where turbulence is weaker. In these regions, the turbulence may not be sufficient to sustain continuous re-acceleration of particles, limiting the production of high-energy electrons and making high-frequency emissions harder to detect with the achieved sensitivity. As a result, radio halos are generally more extended at lower frequencies. The southwestern emission of A384 was not recovered at 1.28 GHz, even at a 2$\sigma$ significance level. Considering the integrated spectral index of the halo, we extrapolate the flux density of the extended region from its 407 MHz measurement and estimate an upper limit of $<0.58$ mJy at 1.28 GHz for any undetected emission in that region. Furthermore, if we consider the 1$\sigma$ detection from the MeerKAT map, we find that the spectrum in this region is very steep, with a slope of $-1.6$. This suggests that the relativistic particle population in this region has aged significantly, reducing the likelihood of detection at higher frequencies.


\section{Dynamical Parameters from X-ray Observations}\label{X-ray results}
To examine the dynamical property of A384 through X-ray morphology, we focus on two morphological parameters: the surface brightness (SB) concentration parameter ($C_\mathrm{{SB}}$) and the centroid shift ($w$) between the X-ray peak and centre of mass. These methods have been used widely in detecting the dynamical status of galaxy clusters \citep{Cassano_2010ApJ, Lovisari_2017ApJ, Yuan_2020MNRAS}.
A study by \citet{Lovisari_2017ApJ} showed that combining centroid shift and $C_\mathrm{{SB}}$ gives the best estimation of cluster dynamical state.

\subsection{X-ray surface Brightness Concentration Parameter}
The ratio of X-ray SB between the core and ambient region or the $C_\mathrm{SB}$ is an excellent method to distinguish between disturbed and relaxed clusters, where a cluster is defined to be relaxed if $C_{\rm{SB}}>$  0.2 and disturbed otherwise \citep{Santos2008A&A}. This method is less suffered by projection effects and thus is very effective even in case of a line-of-sight merger \citep{Cassano_2010ApJ}. For A384, we measured the X-ray flux in two circular apertures of 100 kpc and 500 kpc and thereafter $C_\mathrm{SB}$ is calculated using equation 
\ref{eq4}.
\begin{equation} \label{eq4}
C_\mathrm{SB} = \frac{S < (100\ \mathrm{kpc})}{S < (500 \ \mathrm{kpc})}
\end{equation}
where $S$ is the X-ray surface brightness within a particular radius of 100 kpc and 500 kpc, respectively. The $C_\mathrm{SB}$ was found to be $0.16
\pm 0.01$ for A384 which puts A384 under the classification of dynamically disturbed clusters \citep{Santos2008A&A,Cassano_2010ApJ}.

\subsection{Centroid Shift Measurement}
Cluster merger activities disrupt the central cool core and often shift the X-ray peak from the centre of large-scale X-ray emission or the centre of mass. Thus, measuring the centroid shift is a very effective method to study any dynamical disturbance in the cluster \citep{Poole2006MNRAS}. For measuring the shift for A384, we take consecutive circular apertures starting from 25 kpc with each aperture radius increasing by 5\% from the previous to a maximum aperture radius $R_{\rm{ap}}$ = 500 kpc.
The centroid shift is measured as the standard deviation of the projected distance between the X-ray peak determined from the image and the centre of the X-ray SB obtained within the apertures in units of $R_{\rm{ap}}$ following \citet{Poole2006MNRAS} and using equation \ref{eq5}.

\begin{equation} \label{eq5}
w = \Big[\frac{1}{N-1}\sum(\Delta_i - \left<\Delta \right>)^2\Big]^{1/2} \times \frac{1}{R_\mathrm{ap}}
\end{equation}
where $\Delta_i$ is the distance between the centroid of the i$^{th}$ aperture and the X-ray peak, and $\Delta$ represents the mean of $\Delta_i$. We obtain a centroid shift value of \textit{w} = $0.057 \pm 0.012$ for A384. Previous studies indicate that a cluster can be classified as having a disturbed morphology if the centroid shift, $w > 0.012$ \citep{Cassano_2010ApJ, Weissmann_2013A&A}. Thus, the centroid shift between the X-ray peak and the cluster centre in A384 is well above this threshold, suggesting a disrupted cool core.

\section{Results from Optical Analysis}\label{optical_results}

\begin{figure*}
\begin{center}
\includegraphics[width=\textwidth]{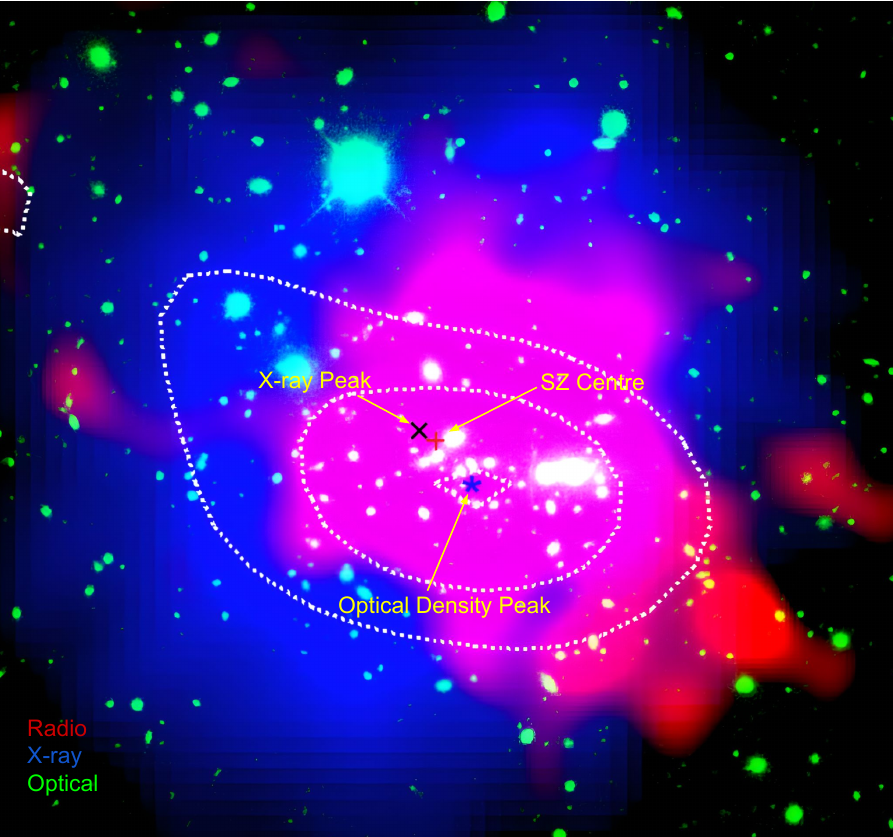}\\
\caption{A multi-wavelength, colour composite image of A384 depicting the radio, X-ray and optical view of the cluster. The RGB image shows uGMRT 400 MHz radio halo emission in red, XMM-Newton X-ray emission in blue and DES optical emission in green. The peak of the optical density map, X-ray map and SZ centre are marked. The dotted line shows the galaxy density distribution in the cluster created using DES observation.}
\label{fig: A384_rgb}
\end{center}
\end{figure*}

Figure \ref{fig: A384_opticaldensity} shows the 2.5 Mpc optical density map composed with each pixel measuring 0.05 Mpc overlaid with GMRT low-resolution radio contours. The optical density map exhibits an elongated structure and a slight offset toward the northeast of the radio halo contours. Moreover, the cluster density peak does not coincide with the X-ray peak or SZ centre (see Figure ~\ref{fig: A384_rgb}). Relaxed clusters often show symmetrical structures with the presence of a BCG at the centre and coinciding peaks of optical density maps with X-ray and SZ peaks. Whereas, an asymmetrical structure or a shift in peaks is the signature of a disturbed cluster morphology \citep{Wen_Han2013MNRAS, Pandge2019MNRAS}. To quantify the dynamical nature of A384, we further considered two tests: centre shift ($CS$) between the cluster's SZ and optical density peak as well as X-ray and optical density peak, and calculation of the asymmetry parameter. 

\subsection{Centre Shift Calculation}
We calculated the offset of the optical density peak from the X-ray peak ($CS_\mathrm{O-X}$) and the SZ peak ($CS_\mathrm{O-SZ}$). The optical, X-ray and SZ peaks are marked in Figure~\ref{fig: A384_opticaldensity}. The offset was calculated using equation \ref{eq10}.
\begin{equation} \label{eq10}
CS=\sqrt{(x_\mathrm{o}-x_\mathrm{x/sz})^2+(y_\mathrm{o}-y_\mathrm{x/sz})^2} \times 0.05~\mathrm{Mpc}
\end{equation}
where x$_\mathrm{o}$ and y$_\mathrm{o}$ are the co-ordinates of the density map peak, x$_\mathrm{x}$ and y$_\mathrm{x}$ are the co-ordinates of the X-ray peak and x$_\mathrm{sz}$ and y$_\mathrm{sz}$ are the co-ordinates of the SZ peak from \citet{Hilton_2018ApJS}.

We obtained $CS_\mathrm{O-X}$  = $0.100 \pm 0.041$ Mpc and $CS_\mathrm{O-SZ}$ = $0.074 \pm0.041$. Calculation of $CS$ has limitations as they are often biased due to projection effects. Thus, we further checked the presence of asymmetry in the cluster.

\subsection{Asymmetry Parameter Calculation}
We calculated the asymmetry parameter for A384 following \citep{Conselice2000ApJ, Wen_Han2013MNRAS}, where we rotated the density map in both vertical and horizontal directions and then subtracted it from the original map. Thereafter, the amplitudes of the pixels of the residual image were compared to the original map for the asymmetry parameter calculation. The asymmetry was calculated following \citet{Wen_Han2013MNRAS} within the R$_{500}$ region of the cluster which is 1.13 Mpc \citep{Piffaretti_2011A&A}. We calculated an asymmetry parameter value $A^2=  0.12 \pm 0.01$ for A384 using equation \ref{eq11}.

\begin{equation} \label{eq11}
A^2 = \frac{\sum(I_\mathrm{0}-I_\mathrm{\phi})^2}{\sum2I_\mathrm{0}^2}
\end{equation}

where $I_\mathrm{0}$ and $I_\mathrm{\phi}$ represents the original and rotated image, respectively. The measure of $I_\mathrm{0}-I_\mathrm{\phi} = 0$ signifies symmetry, and 1 signifies complete asymmetry. 


\section{Discussion}\label{discussion}
Interpretation of observations of merging clusters is difficult as the observations are often biased due to the projection effects. Radio halos are mostly found in merging clusters, and the presence of the $\sim$700 kpc radio halo in Abell 384 itself suggests the cluster is in a merging state. However, recent studies have shown the presence of radio halos in relaxed clusters \citep{Bonafede_2014MNRAS, Savini2018MNRAS}. Therefore, a comprehensive study of the cluster's dynamical state is essential for a complete understanding of the origin of radio halos. Here, we discuss the dynamical nature of A384 and the possible origin of the radio halo in the cluster.

\subsection{Dynamical nature of A384}
The integrated spectral index obtained for the radio halo in Abell 384 is slightly flatter than typically observed halos. Moreover, the spectral index distribution exhibits a non-uniform, clumpy structure. This can be explained by the turbulent re-acceleration model, where turbulence created during cluster mergers cascades to smaller scales in the ICM, and the varying acceleration efficiency across different regions of the ICM can lead to this type of non-uniform spectral index distribution.

From the X-ray analysis, we estimated the concentration parameter, $C_{\rm{SB}} = 0.16 \pm 0.013$ and the centroid shift value, $w = 0.057 \pm 0.012$. 
It is important to note that the 
centroid shift method can be influenced by mergers occurring along the line of sight, whereas the concentration parameter is not sensitive to projection effects and, therefore, provides a more robust result. For A384, the low $C_{\rm{SB}}$ value of 0.016, which is significantly below the dynamical state threshold of $C_{\rm{SB}} \geq 0.2$, and the larger centroid shift of X-ray peak from cluster centre both indicates the merging state of the cluster.

\begin{figure*}
\includegraphics[width=0.45\textwidth]{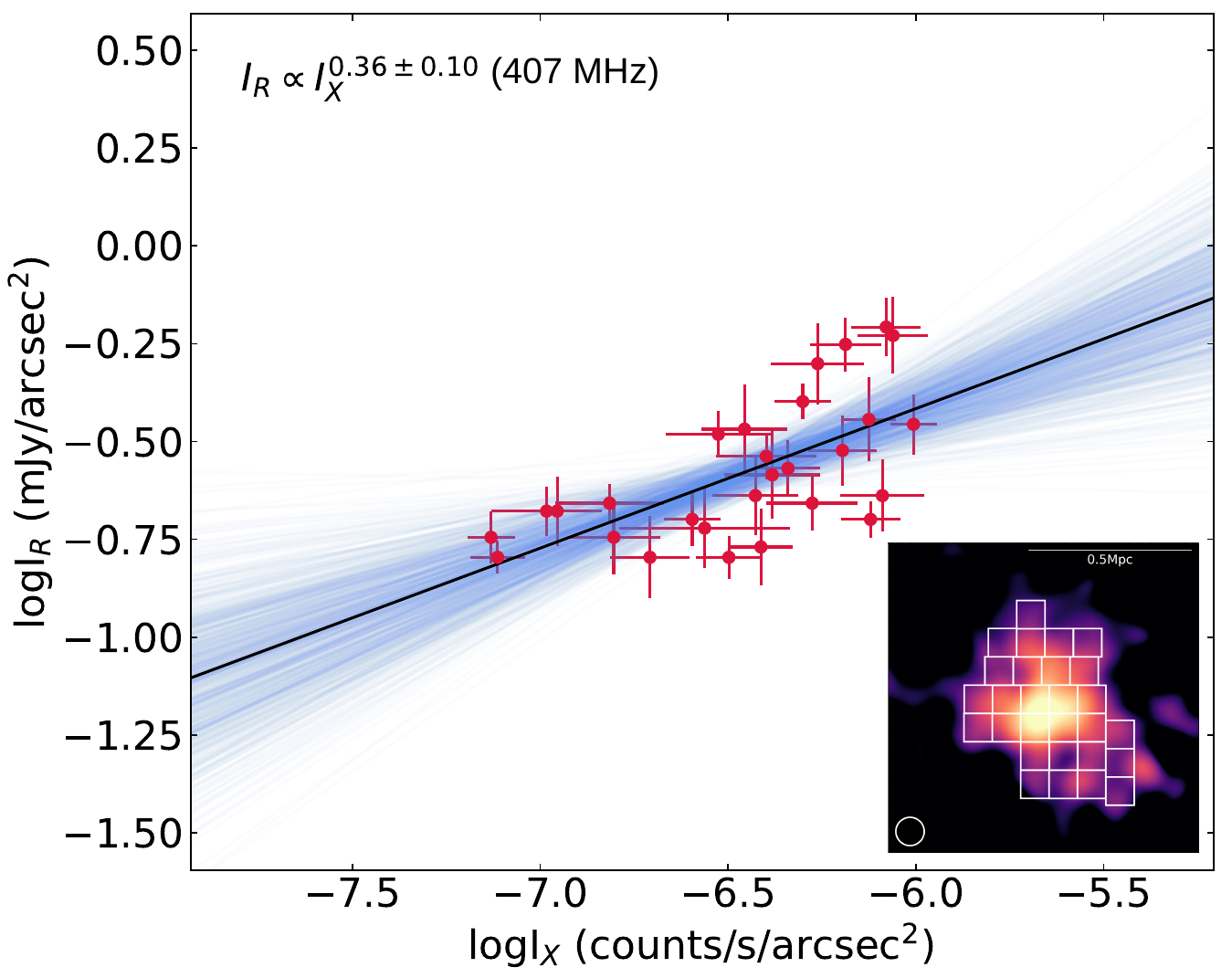}
\includegraphics[width=0.46\textwidth]{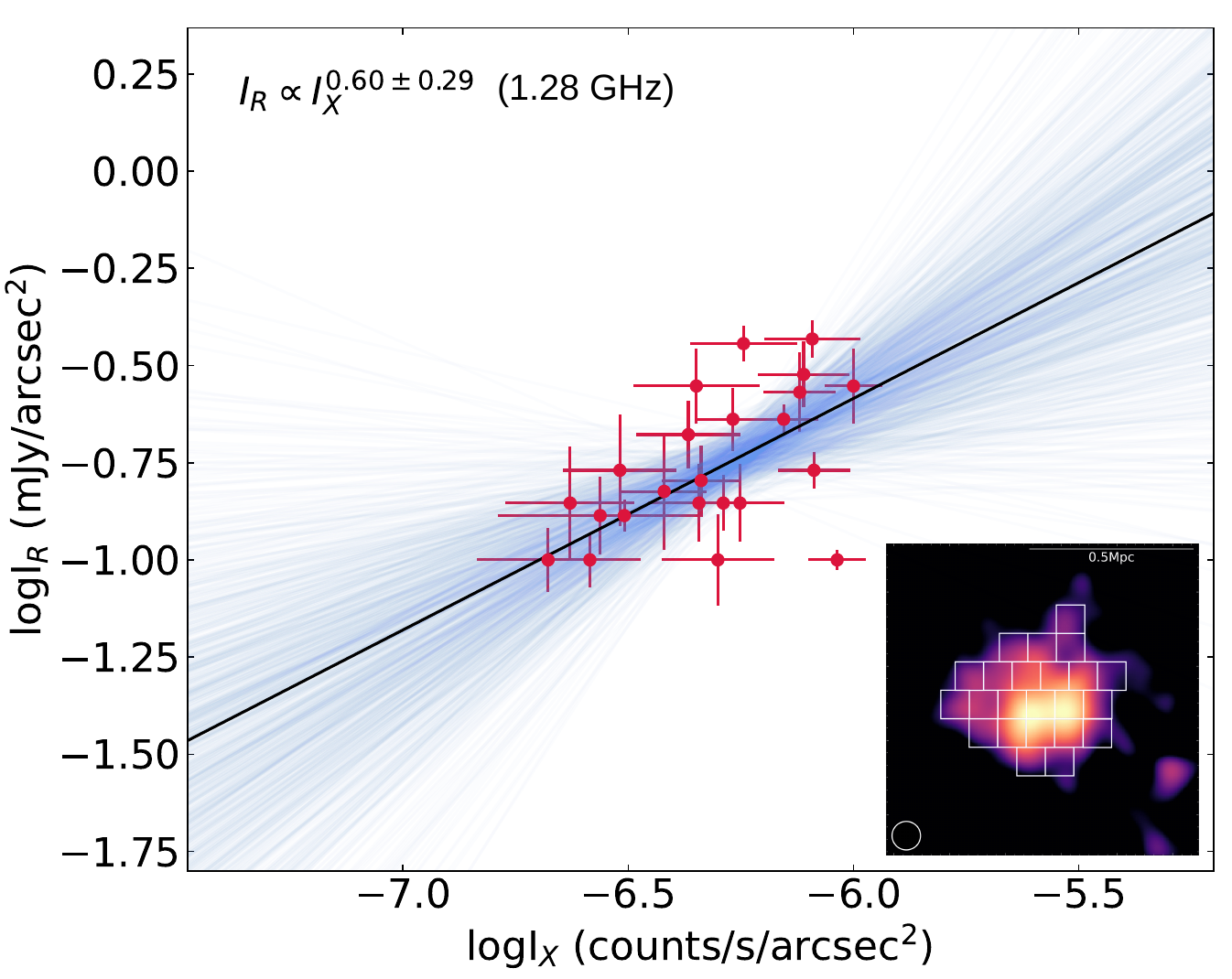}
\caption{Left: Radio (407 MHz) and X-ray surface brightness correlation of the radio halo in A384, extracted in square boxes of 23$\arcsec$. Right: The correlation study between 1.28 GHz radio and X-ray image. The respective radio images overlaid with boxes are shown at the bottom right corners. The black lines indicate the best-fit line obtained with \textsc{LINMIX}, and the blue lines represent the MCMC posterior chains.}
\label{fig: A384_linmix_400}
\end{figure*}

The optical observation of the cluster reveals an elongated galaxy density map hinting at the potential merger axis along the northeast-southwest direction. The measured $CS_\mathrm{O-X}$ is $0.100 \pm 0.041$ Mpc and $CS_\mathrm{O-SZ}$ is $0.074 \pm 0.041$ Mpc. Analysing a set of 98 clusters, \citet{Pillay_2021Galaxy} showed that while clusters with CS$_\mathrm{O-SZ}>$ 0.4 were all disturbed, both relaxed and disturbed systems could exhibit $CS_\mathrm{O-SZ}$ values below this threshold. The primary reason for this overlap below 0.4 Mpc is that $CS$ measurements are sensitive to projection effects. For instance, a relaxed cluster may harbour small substructures or experience minor interactions that produce small offsets, resulting in a $CS_\mathrm{O-SZ}<$ 0.4 Mpc. Conversely, a disturbed cluster observed along a specific line of sight may appear more symmetric, yielding a smaller $CS_\mathrm{O-SZ}$. Thus, in the case of A384, the $CS_\mathrm{O-SZ}$ value of 0.074 Mpc alone does not eliminate the possibility of dynamical disturbances and underlying substructures in the cluster. On the other hand, $CS_\mathrm{O-X}$ values exceeding 0.05 R$_{500}$ have been demonstrated to indicate dynamical disturbances within galaxy clusters \citep{Ota_2023A&A, Seppi_2023A&A}. For A384, the measured $CS_\mathrm{O-X}$ of 100 kpc significantly exceeds the corresponding threshold of 0.05 R$_{500}$ = 57 kpc,  indicating dynamical activity in the cluster.
The asymmetry parameter estimated for A384 is $A^2 = 0.12 \pm 0.01$. \citet{Pillay_2021Galaxy} observed that clusters with an asymmetry $A^2 > 0.1$ could be typically classified as dynamically disturbed, regardless of their CS values. Thus, the asymmetry parameter provides additional evidence that A384 is a disturbed cluster. We note that this study relies on a 2-D map of the cluster, which limits our ability to completely understand line-of-sight disturbances. A more comprehensive approach to studying the merging geometry would involve the use of spectroscopic redshift distribution and velocity measurements of cluster galaxies \citep{Girardi_2016MNRAS, Barrena_2022A&A}.

\subsection{Radio, X-ray Correlation Study and Origin of the Halo:}
The similarity of the morphology of radio halos with the cluster X-ray emission, as observed in several cases, indicates a connection between the thermal non-thermal components (see \citealt{Feretti_2012A&A,vanweeren_2019SSRv}. In the case of A384, the radio halo is observed to be slightly offset from the cluster's X-ray map. Thus, to further understand the origin of the halo, we perform the point-to-point correlation study between radio halo and X-ray emission. 
This is an effective method to understand the spatial variation of relativistic particles and magnetic fields with ICM gas. Previous studies on radio halos have shown that halo emissions show a robust correlation with X-ray emission \citep{Govoni_2001A&A,Rajpurohit_2021A&A, Sikhosana_2023MNRAS, Balboni_2024A&A, Santra_2024ApJ}. The radio and X-ray surface brightness relation can be expressed as a power law  
\begin{equation}
\log\left(\frac{I_{\mathrm{R}}}{\text{mJy/arcsec}^2}\right) = a + b ~\log\left(\frac{I_{\mathrm{X}}}{\text{counts/s/arcsec}^2}\right)
\end{equation}
where $I_{\rm{R}}$ and $I_{\rm{X}}$ are the radio and xray surface brightness respectively. A linear correlation ($b = 1$) indicates that the energy density of relativistic electrons and magnetic field strength in the ICM scales proportionally with the energy density of thermal particles. A sub-linear correlation ($b<$1) implies a slower spatial change of non-thermal components compared to thermal components. Conversely, a super-linear correlation ($b>$1) implies a faster spatial change in radio emission relative to X-ray emission \citep{Govoni_2001A&A, Balboni_2024A&A}.
In the case of turbulent acceleration during cluster mergers, a small fraction of energy is cascaded to accelerating particles to relativistic speed and amplification of magnetic field while the rest of the energy is converted into thermal energy in ICM. Thus, a sublinear correlation is expected if the radio halos originated from the turbulent re-acceleration in ICM \citep{Govoni_2001A&A}. 

To investigate the thermal-nonthermal interplay in A384, we performed a point-to-point correlation study employing square grids of 23$\arcsec$, considering the image resolution, across the radio halo. The grids were placed in regions having surface brightness exceeding 3$\sigma$ in the radio maps of 407 MHz and 1.28 GHz, and the power-law fitting was conducted using the \textsc{LINMIX} package\footnote{\url{https://linmix.readthedocs.io/en/latest/src/linmix.html}} \citep{Kelly_2007ApJ}. \textsc{LINMIX} performs Bayesian linear regression, drawing output parameters from the posterior distribution through Markov Chain Monte Carlo (MCMC) sampling. The observed surface brightness profiles of the A384 radio halo follow a positive sub-linear correlation with a slope $b_{407~\mathrm{MHz}} = 0.36 \pm 0.10$ at 407 MHz and $b_{1.28~\mathrm{GHz}} = 0.60 \pm 0.29$ at 1.28 GHz. The best-fit parameters and the Spearman ($r_{\rm{s}}$) and Pearson coefficients ($r_{\rm{p}}$) for the fit are mentioned in Table \ref{tab: correlation}. The slopes show a frequency dependence, with the steeper slope at higher frequencies, which is expected since high-frequency electrons lose energy faster than low-frequency electrons. A similar trend has been observed in the radio halo of MACS J0717.5+3745 \citep{Rajpurohit_2021A&A_macs_halo}. The $I_{\rm{R}}$-$I_{\rm{X}}$  sub-linear trend indicates that higher X-ray luminous regions or the denser regions fail to produce a proportionally higher radio emission. Such can happen in case of turbulent acceleration during cluster merger, where the turbulence energy dissipates more via collisionless damping at denser regions reducing the particle acceleration efficiency \citep{Brunetti_2007MNRAS, Pinzke_2017MNRAS}. Thus the correlation trend in A384 provides evidence for a turbulence-driven mechanism behind the radio halo's origin.

\begin{table}
    \centering
    \caption{The best-fit parameters for point-to-point correlation study along with Spearman ($r_\mathrm{s}$) and Pearson ($r_\mathrm{p}$) correlation coefficient obtained for the radio halo at 407 MHz and 1.28 GHz.}
    \begin{tabular}{cccc}
        \hline
        Frequency & Slope & \( r_\mathrm{s} \) & \( r_\mathrm{p} \) \\
        \hline
        407 MHz  & $0.36\pm0.10$  & 0.70 & 0.64 \\
        1.28 GHz  & $0.60\pm0.29$  & 0.53 & 0.54 \\
        \hline
    \end{tabular}
    \label{tab: correlation}
\end{table}

\section{Summary and Conclusion} \label{summary}
A384 is a massive galaxy cluster identified to be hosting a radio halo by \citetalias{Knowles_2021MNRAS} during the MERGHERS pilot survey. We conducted a comprehensive multi-wavelength study on A384 to investigate the dynamical state of the cluster and the origin of the radio halo. The key findings from our radio, X-ray, and optical observations are summarized below:
\begin{enumerate}
    \item \textbf{Radio spectral Characteristics of the halo:}
  \begin{itemize}
    \item We detected an extensive $\sim$700 kpc radio halo at 407 MHz in A384, tracing the southwestern region of the thermal X-ray emission. Additionally, utilizing legacy GMRT observations, we observed the halo extending approximately 460 kpc at 608 MHz.
    \item The radio halo exhibits a relatively flatter spectral index between 407 MHz and 1.28 GHz ($\alpha^{1284}_{407} = -0.96 \pm 0.04$), indicative of recent particle acceleration events likely triggered by merger activities. The spectral index map reveals a patchy structure, suggesting a complex particle acceleration history.
\end{itemize}

    \item \textbf{X-ray Observations:}\\
The X-ray map of A384 shows a significant offset between the X-ray peak and the emission centre, with a centroid shift of $w = 0.057$, signalling a disturbed ICM. This disturbance is further corroborated by the low $C_{\rm{SB}} = 0.016$.

    \item \textbf{Optical Observations:}\\
The optical galaxy density map reveals an elongated, axisymmetric structure along the northeastern and southwestern directions, hinting at the potential merger axis. The radio contours align with the optical galaxy density distribution. The calculated offset between optical and X-ray peaks and asymmetry parameter further support the presence of a dynamically disturbed cluster environment. 
\end{enumerate}

The radio halo observed in A384 differs from typical textbook examples, as its emission is displaced from the cluster’s X-ray distribution. Nevertheless, the patchy spectral index distribution and the dynamical disturbances seen in X-ray and optical data support the turbulent reacceleration theory, aligning with observations of other halos (see \citealt{vanweeren_2019SSRv} for a review). Notably, no particular radial steepening was observed in this case. The sublinear correlation between the radio and X-ray surface brightness further strengthens the idea that the radio halo’s origin is primarily driven by turbulent reacceleration mechanisms within the ICM. 

While our study provides some important findings using 2-D maps of the cluster, it does have limitations in capturing the complexities of line-of-sight disturbances. Moreover, in the number of clusters where radio-X-ray correlation study has been performed over a wide range of frequencies, a difference in correlation trend has been observed. In a few clusters, the sub-linear correlation slope has been observed to remain constant over multiple frequencies \citep{Rajpurohit_2021A&A, Santra_2024ApJ}; whereas, in our case, it changes with frequencies like in the cases for \citet{Hoang_2019A&A_A520, Rajpurohit_2021A&A_macs_halo, Balboni_2024A&A}. A linear trend has been observed in the radio halo for A2255 \citep{Govoni_2001A&A, Botteon_2020ApJ}. Thus, the information from a handful of clusters indicates that the thermal-nonthermal interconnection in ICM is much more complex. Consequently, further studies of this kind are needed to gain a more comprehensive understanding of the processes behind these phenomena and the genesis of radio halos in galaxy clusters.

\section*{Acknowledgements}
We thank the anonymous reviewer for the comments and suggestions. We would like to thank IIT Indore and Rhodes University for giving out the opportunity to carry out the research project. We acknowledge the
funding via the Department of Science and Technology, Government of India sponsored DST-FIST grant no.
SR/FST/PSII/2021/162 (C). SC acknowledges the funding support by the National Research Foundation (NRF). RR's research is supported by the South African Research Chairs Initiative of the Department of Science and Technology and the NRF (Grant id. 81737). KK acknowledges funding support from NRF-SARAO (UID95492). MR acknowledges support from the National Science
and Technology Council (NSTC) of Taiwan (NSTC 112-2628-M-007-003-MY3).
SPS acknowledges the financial assistance of the SARAO. We thank the staff of the GMRT and MeerKAT who have made the radio observations possible. The GMRT is run by the National Centre for Radio Astrophysics of the Tata Institute of Fundamental Research. The MeerKAT telescope is operated by the South African Radio Astronomy Observatory, which is a facility of the National Research Foundation, an agency of the Department of Science and Innovation. This research has used data obtained from the XMM Newton Science Archive and the Dark Energy Survey Archive. This research made use of Astropy,\footnote{http://www.astropy.org} a community-developed core Python package for Astronomy \citep{Astropy_2013A&A...558A..33A,astropy_2018}, matplotlib\citep{Matplotlib_Hunter:2007}, and APLpy, an open-source plotting package for Python \citep{APLpy_2012ascl.soft08017R}. 

\section*{Data Availability}
The radio data and images underlying this article will be shared on reasonable request to the corresponding author. Moreover, the data used in our work are available in the GMRT online archive (\url{https://naps.ncra.tifr.res.in/goa/data/search}), the MeerKAT data archive (\url{https://apps.sarao.ac.za/katpaws/archive-search} and the XMM science archive (\url{http://nxsa.esac.esa.int/nxsa-web/#search})



\bibliographystyle{mnras}
\bibliography{example} 




\appendix

\section{verification of spectral index distribution}\label{Appendix}
\begin{figure*}
\includegraphics[width=\columnwidth]{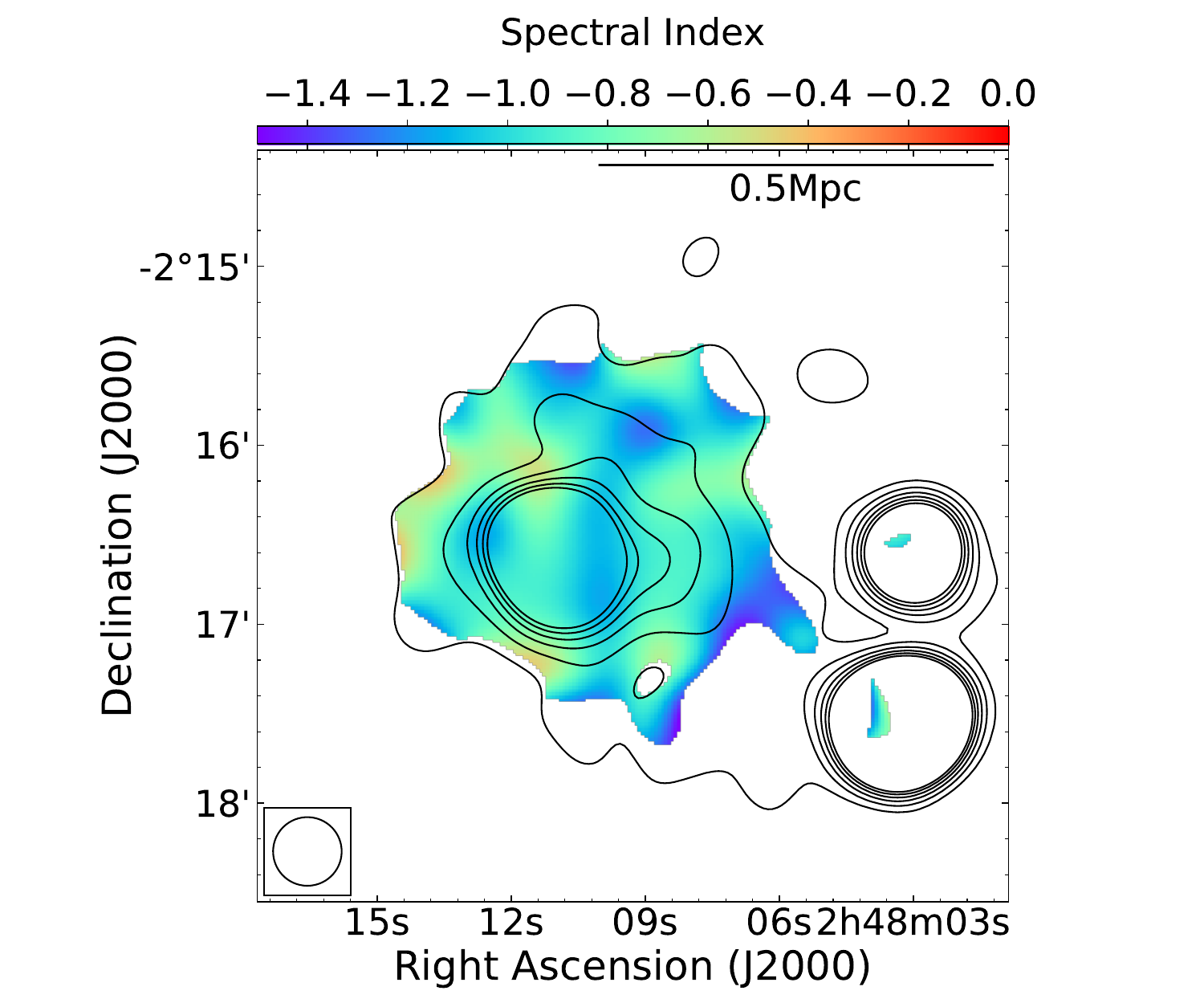}
\includegraphics[width=\columnwidth]{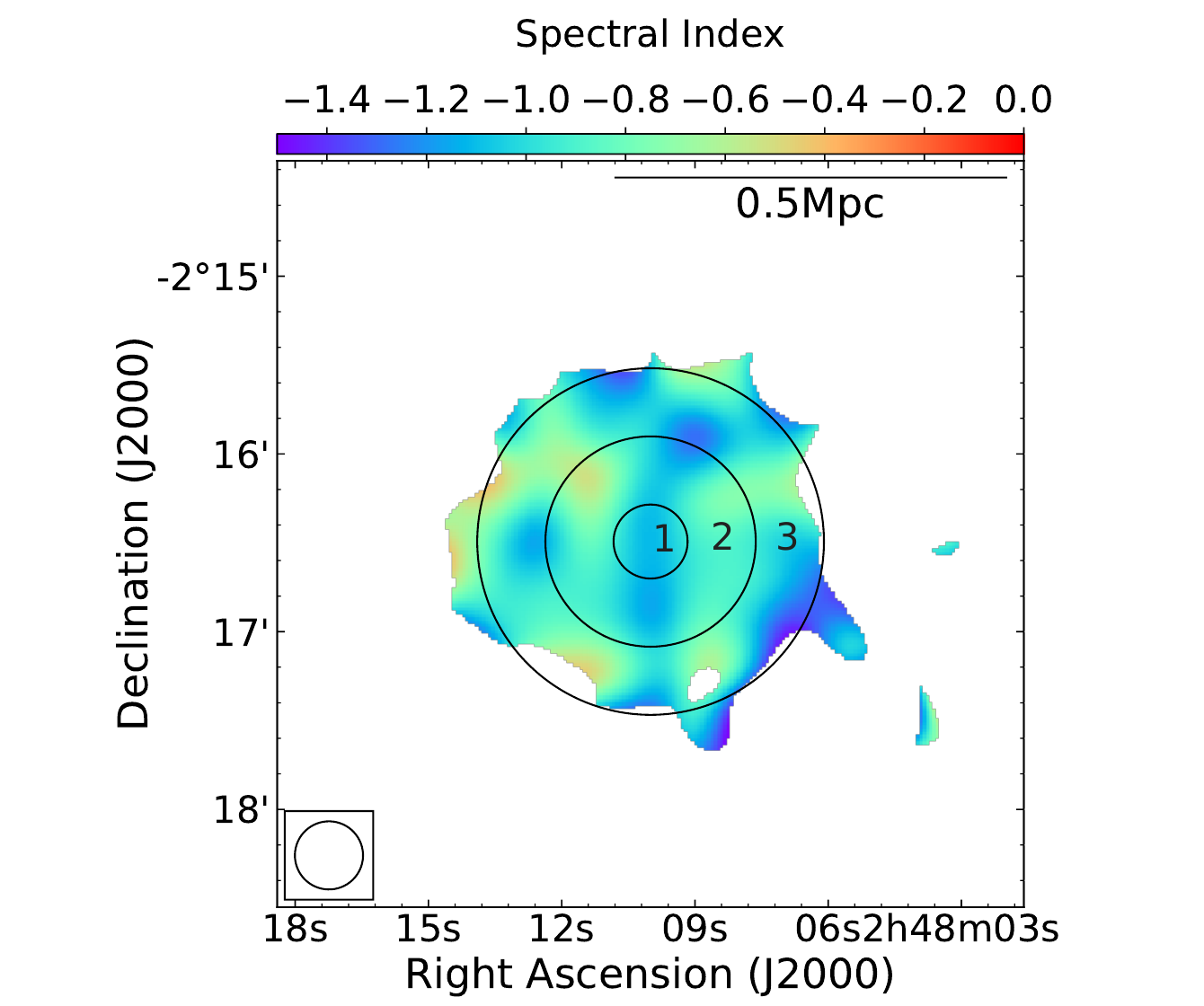}
\caption{Left: Spectral index map of A384 overlaid with low-resolution (restoring beam $23\arcsec$) uGMRT 407 MHz radio contours without subtraction of compact sources. The contour are placed at (1,2,3,4,5,6)$\times3\sigma_\mathrm{407 ~MHz}$, where $\sigma_\mathrm{407 ~MHz}= 110\ \mu$Jy/beam. Right: Spectral index map of A384 overlaid with annuli of width 23$\arcsec$.}
\label{fig: A384_spix_appendix}
\end{figure*}

We verified the spectral index distribution in the cluster and whether the map was affected by compact source subtraction. Figure~\ref{fig: A384_spix_appendix}(Left) shows the spectral index map of A384 overlaid with the 407 MHz low-resolution contours of the cluster radio emission without subtracting the compact sources. The map shows that the variation in the spectral index distribution throughout the halo emission is independent of region, including or excluding compact sources. 

Furthermore, we created circular annuli with a width of 23 $\arcsec$ across the halo to check for any radial steepening in the integrated spectral index from cluster centre to outskirt (Figure~\ref{fig: A384_spix_appendix}, Right). There was no particular radial steepening observed, and the obtained median spectral indices are $\alpha_{\rm{reg1}} = -1.07\pm 0.12$ in region 1, $\alpha_{\rm{reg2}} = -0.92\pm 0.15$ in region 2 and $\alpha_{\rm{reg3}} = -0.93\pm 0.25$ in region 3.



\bsp	
\label{lastpage}
\end{document}